\documentclass[a4paper,11pt]{article}
\usepackage{graphicx}
\usepackage[space]{grffile}
\usepackage{latexsym}
\usepackage{textcomp}
\usepackage[normalem]{ulem}
\usepackage{colortbl}
\usepackage{amsfonts,amsmath,amssymb}
\usepackage{natbib}
\usepackage[hidelinks]{hyperref}
\usepackage{subcaption}
\usepackage{etoolbox}
\makeatletter
\usepackage{authblk}
\usepackage[utf8]{inputenc}
\usepackage[english]{babel}
\usepackage[top=0.8in, bottom=0.8in, left=0.8in, right=0.8in]{geometry}

\title{\textbf{Northward Propagating Versus Non-propagating BSISO over South Asia: Horizontal Advection Driven Moisture Mode Within a Vertically Sheared Background}}

\author[1,2,*]{Sambrita Ghatak \thanks{Corresponding author: Sambrita Ghatak (pulu.dec@gmail.com)}} 
\author[1,2,*]{Jai Sukhatme} 

\affil[1]{Centre for Atmospheric and Oceanic Sciences, Indian Institute of Science, Bangalore, India.}
\affil[2]{Divecha Centre for Climate Change, Indian Institute of Science, Bangalore, India.}
\affil[*]{These authors contributed equally to this work.} 

\begin{document}
	
\maketitle
\begin{abstract}
    \normalsize{The Boreal Summer Intraseasonal Oscillation (BSISO) is a pronounced mode of tropical variability. Here, we identify two types of BSISO events, one which propagates northward over South Asia (SA) from the equatorial Indian Ocean (EIO), and the other which doesn't. Contrasting their behaviour shows that northward propagation occurs in multiple stages after convection is initiated over the EIO. First, convection moves into the southern Arabian Sea (AS) due to moistening of the free troposphere via horizontal BSISO anomalous winds acting on the background moisture distribution, and forms a northwest-southeast (NW-SE) oriented convection band. Subsequently, in the presence of an easterly vertical shear of monsoon winds and meridional gradient of anomalous vertical velocity, a NW-SE oriented tilting term is generated that results in a tilted gyre north of the existing convective anomaly and south-easterly BSISO winds over the South Asian landmass. In the second stage, these winds tap the ambient north-westward moisture gradient and help move convection further north over land. Moreover, background winds advect anomalous moisture to initiate convection over the Bay of Bengal. For non-propagating events, though a Rossby gyre results as a response to nascent EIO convection, it is smaller, thus BSISO advection of moisture is weaker and does not initiate convection over the southern AS. In turn, the meridional gradient of anomalous vertical velocity is weak, and the background vertical shear does not generate sufficient tilting over the northern AS. Thus, the convective wind response stalls, and large-scale convection does not propagate north of 15N. Thus, free-tropospheric moisture advection and vortex tilting due to the background vertical shear  work together for robust northward propagation of the BSISO.}
\end{abstract}

\section{Introduction}

\noindent Intraseasonal oscillations (ISOs) in the tropical atmosphere exhibit pronounced seasonality \citep{wang1990synoptic,wheeler2004all}. While the Madden Julian Oscillation (MJO) is the dominant ISO signal during winter, the Boreal Summer Intraseasonal Oscillation (BSISO) is the most significant intraseasonal signal during the northern hemisphere summer, particularly in the Indo-Pacific sector. 
The MJO and BSISO develop in the Indian Ocean and have similar timescales \citep{wheeler2004all,kiladis2014comparison,wang2018propagation}, i.e., both have a time period roughly between 30-60 days, but they differ markedly in their spatial pattern and propagation characteristics. While the MJO is largely symmetric about the equator and predominantly characterized by equatorial eastward propagation, the most prominent feature of the BSISO is dramatic northward propagation from the equatorial region to South Asia (SA) and the western North Pacific (WNP) Ocean, though it has a weaker eastward propagating component also \citep{lawrence2002boreal,lee2013real,adames2016seasonality,wang2018propagation,li2018northward,singh2020tracking}. 
Historically, the MJO and BSISO have been seen as separate low-frequency ISO modes with similar time scales, but the conceptual boundary between them is not very clear \citep{wang2022unified}, and a few recent studies don't see the BSISO and MJO as separate phenomena \citep{adames2016seasonality,jiang2018unified,wang2022unified}, rather the BSISO is viewed as a ``northern summer incarnation of the MJO" \citep{wang2022unified}.

\noindent Though northward propagation of the BSISO is observed over SA as well as in the WNP, in this paper, we focus on SA. Over the South Asian monsoon region, large-scale northward propagating convective activity with a period of around 40 days was first observed by \cite{yasunari1979cloudiness} and \cite{sikka1980maximum}, though the existence of  a mode with such a timescale over this region was noted much earlier --- see the discussion in \cite{li2018paper}. Since then, it has been recognized that this envelope of coherent convection, or the BSISO, has a profound influence on regional weather systems and extremes, such as  monsoon active and break cycles \citep{krishnamurti198230,rajeevan2010active,pai2011impact,prasanna2012moist,dey2022intraseasonal}, monsoon onset and retreat \citep{moron2012impact,taraphdar2018mjo,karmakar2019relation,xie2022mechanism}, droughts 
\citep{bhat2006indian,neena2011leading}, floods \citep{singh2024calibrated}, 
tropical cyclones \citep{camargo2009diagnosis,kikuchi2010formation} and monsoon low-pressure systems \citep{goswami2003clustering,krishnamurthy2010composite,nikumbh2021multiscale,karmakar2021influence,hunt2022non}. 
Thus, the BSISO  strongly influences the overall pattern and amount of of monsoon rainfall by influencing its important components \citep{goswami2001intraseasonal,goswami2005dynamics}. 
Representation of the BSISO still remains a major challenge for regional and global climate models \citep{neena2017model,ghosh2023signals,xiang2024prediction}, which adversely affects overall monsoon prediction \citep{keane2021biases,joseph2023evaluation}. 
Due to its impact, a lot of attempts have been made to understand the BSISO \citep{kikuchi2021boreal}. But, despite progress, the system remains elusive, and particularly the mechanism behind its striking northward movement is not clearly understood \citep{kikuchi2021boreal,wang2022unified}. 
In fact, in a recent review, \cite{bellon2023selected} documented the BSISO as an outstanding puzzle in tropical atmospheric variability. 

\noindent   
While early studies suggested varied mechanisms \citep{webster1983mechanisms,goswami1984quasi,nanjundiah1992intraseasonal,wang1997model}, broadly, there are two dominant modern schools of thoughts that attempt to explain northward propagation of the BSISO  \citep{jiang2018unified,wang2020diagnosing}. The older amongst these understood the BSISO in terms of the internal dynamics of the atmosphere using simplified models, where interaction between the circulation generated by convection and the monsoon background flow is the key \citep{jiang2004structures,drbohlav2005mechanism,bellon2008instability}. There are variations within this school of thought, but the basic understanding of northward propagation of convection is via generation of free-tropospheric barotropic cyclonic vorticity to the north of the existing convection, which causes boundary layer moisture convergence (BLMC) resulting in a northward shift in convection. 

\noindent Though widely used, these mechanisms are subject to growing skepticism with the advent of the moisture mode framework. This framework stands upon the observation that tropical convection is primarily dictated by column-moisture \citep{bretherton2004relationships}, and related theoretical analysis which showed that completely new modes can emerge when prognostic moisture is included in dynamical models \citep{sobel2001weak}. It has been shown that prognostic moisture is essential to produce MJO-like propagating modes in simpler models as well as idealized general circulation models \citep{suhas2022}. According to this view, as MJO moist convection follows anomalous column-integrated water vapor \citep{kiranmayi2011intraseasonal,kim2014propagating,adames2015three}, to understand the eastward propagation of the MJO, the evolution of column-moisture is studied in the form of moisture/moist static energy(MSE)/moist entropy(ME) budget. There are variations within this framework based on the process of moistening. One idea, popularly known as WISHE (wind-induced surface heat exchange)-moisture mode, is where wind induced evaporation helps in moistening to the east of the convection \citep{fuchs2005large,fuchs2017simple,khairoutdinov2018intraseasonal}, though observational evidence does not appear to support this view of the MJO \citep{benedict2007observed,kiranmayi2011intraseasonal, de2015mjo}. The most popular view stresses on the primary importance of horizontal moisture advection \citep{sukhatme2014low,sobel2014moist,adames2016}, though some importance of vertical advection has also been noted \citep{adames2015three,hsu2012role}. The exact process of horizontal advection is being debated within this school of thought \citep{kim2014propagating}. Following observational evidence \citep{kim2014propagating}, and theoretical studies \citep{sukhatme2014low,adames2016}, where an MJO-like eastward propagating mode crucially depends on the background moisture gradient, importance of horizontal advection of background moisture/MSE/ME in the lower free-troposphere by MJO perturbation winds  behind MJO eastward propagation has been established by observational, modelling and theoretical efforts \citep{nasuno2015moistening,kim2017does,gonzalez2017winter,jiang2017key,kim2017impact,kang2020role,kang2021role,Ahmed2021} --- see also the recent reviews by \cite{zhang2020four,jiang2020fifty}.

\noindent Recently, a similar approach has been adopted to understand the BSISO \citep{ajayamohan2011poleward,wong2011apparent,pillai2016moisture,adames2016seasonality,jiang2018unified,wang2020diagnosing,gao2019diagnosing,wang2022unified}. In fact, it has been suggested that both BSISO and MJO can be understood under a unified moisture mode framework, where horizontal advection of background moisture by ISO winds is the key behind propagation for both the MJO and BSISO \citep{jiang2018unified,wang2022unified}. Indeed, the influence of background moisture distribution on intraseasonal modes had been noted in simplified moist models \citep{sobel2001weak,sukhatme2014low,joyQG,adames2016,suhas2020moist,Ahmed2021}, following which \cite{wang2022unified} put forth a unified moisture mode theory, in which they proposed a simple model which can produce both eastward and northward propagating ISOs depending on the imposed background moisture gradient which mimics seasonally changing patterns. 

\noindent  If the moisture mode is defined broadly (that the convection is primarily dictated by moisture anomalies, and prognostic moisture is needed to capture the mode), another school of thought has recently emerged with the introduction of a new class of idealized models \citep{liu2017effects,wang2016trio,wang2017general,liu2016role,chen2019dynamic}.  These include prognostic moisture, just like the traditional moisture mode, but in these models, active feedback between moisture, convective heating and  boundary layer dynamics is critical, as the crucial process of moistening ahead of the convection is BLMC, unlike the traditional moisture mode view described above where horizontal advection is of primary importance for moistening ahead of convection (see also \cite{hu2021reexamining}). This theory can be viewed as an update to the original frictionally-coupled Kelvin-Rossby mechanism \citep{wang1990dynamics,wang1994convective,hendon1994life,maloney1998frictional}, which didn't include prognostic moisture. 
Here, we call this the BLMC moisture mode. 
As nicely explained in \cite{wang2020diagnosing}, although the BLMC view started outside the purview of the moisture mode framework in the context of BSISO, just like MJO, BLMC theory can be treated under  moisture mode framework as BLMC can be viewed as a process of moistening in the column, which causes the propagation of the convection. To summarize, under broader moisture mode framework, there are two dominant viewpoints in context of the BSISO, one where moistening to the north of the convection centre occurs via BLMC, and the second where horizontal moisture advection is primarily responsible for the moistening and propagation of convection. 

\noindent Notably, another mechanism that has been proposed for northward propagation is air-sea interaction \citep{kemball2001equatorial,roxy2007role,bellon2008ocean,sharmila2013role,zhang2018role}, which developed following the observation that warm SST anomalies are present to the north of the active BSISO convection \citep{sengupta2001oscillations,sengupta2001coherent,vecchi2002monsoon,vialard2012processes}. In fact, the effect of air-sea interaction at intraseasonal scale can also be incorporated into the moisture mode framework, as ocean can help in column-moistening by enhancing surface evaporation, i.e., WISHE  \citep{emanuel1987air,neelin1987evaporation} and/or by BLMC \citep{lindzen1987role,back2009relationship}.

\noindent While there are many views summarized above, the mechanisms proposed are far from conclusive, and the lack of consensus in the literature limits our understanding of BSISO \citep{kikuchi2021boreal}. Though the proposed mechanisms have been successful to some extent to explain observations, many fundamental features associated with the BSISO still remain unexplained, such as the evolution of convection and circulation away from the equator (mostly over land), the characteristic northwest-southeast (NW-SE) tilt of the convection band and the role of the background vertical shear of the monsoon winds. 





\noindent In this paper, inspired by the work on the MJO \citep{kim2014propagating}, we try to find the key mechanisms behind northward propagation of BSISO convection by identifying two types of BSISO events, one set where the convection moves northward over SA from the Equatorial Indian Ocean (EIO), and the other where convection does not propagate northward in spite of a strong start over the EIO. Indeed, there has been a characterization of different types of BSISO events, but this has mainly focused on the state of the oceans\citep{ajayamohan2008influence,kottapalli2022weakening,chen2020influences,lee2022dominant}. Further, studies on the diversity of the BSISO \citep{pillai2016moisture,chen2021diversity,strnad2023propagation} have classified both northward and eastward motion of it over the broader Asian Monsoon region. In our identification, we only use poleward movement of coherent convective activity over SA and the difference between the dynamics of these two categories allows us to identify the critical mechanisms behind the northward propagation in this region. Specifically, we employ a moisture mode framework, investigate the moistening process, and show which type of moisture mode view is best suited to understand the BSISO.  We also carry out vorticity budget analysis to understand the evolution of the circulation response and its coupling with convection. Section 2 contains a description of the data used and Section 3 describes the northward propagating and non-propagating BSISO events. We then proceed to the analysis of moisture and vorticity budgets in Sections 4 and 5, respectively. Section 6 discusses our findings in the context of previous theories of the BSISO and conclusions are presented in Section 7.

\section{Data and Methodology} 
Daily meteorological fields from the ERA5 reanalysis project serve as the main dataset for this study \citep{hersbach2020era5}. 
Specifically, we have used 30 years (1980--2009) of horizontal winds, vertical (pressure) velocity, and specific humidity data at 17 pressure levels (1000 to 200 hPa) with an interval of 50 hPa. The horizontal resolution of the data used is 2.5$^\circ$, 
which is used to calculate the derived fields presented in this paper. Daily, 2.5$^\circ$ horizontal resolution outgoing longwave radiation (OLR) data from the National Oceanic and Atmospheric Administration (NOAA) satellites serves as a proxy for moist tropical convection \citep{liebmann1996description}. 

\noindent To isolate the BSISO signal, we used a lanczos filter \citep{duchon1979lanczos} with a 25-70 day bandwidth. There are many studies which used 20-70 days filter to isolate BSISO \citep{yang2019mechanisms,wang2018propagation,liu2022intraseasonal}, also wider ranges such as 20-90 or 20-100 have been used \citep{jiang2018unified,zhang2018role}. As there is a prominent 10-20 day quasi-biweekly mode in this region in the boreal summer \citep{chatterjee2004structure}, and this region is also characterised by a strong seasonal cycle, we use 25-70 days to avoid any overlap with the higher and lower-frequency end, though slightly changing the filter cutoff doesn't significantly affect our results. Prior to filtering, the annual cycles for individual years are removed. The annual cycle is constructed by adding the mean and the first three Fourier harmonics. The composite mean of this annual cycles during the selected BSISO events is called the seasonal background signal in this paper.  

\noindent To distinguish between propagating and non-propagating BSISO events, we used two reference boxes, one over EIO (70$^\circ$-90$^\circ$E, 0$^\circ$-5$^\circ$N), and another over South Asian landmass (70$^\circ$-90$^\circ$E, 17.5$^\circ$-22.5$^\circ$N). The standard deviation of box-averaged OLR in the EIO is $\sim$ 17.2 W m$^{-2}$ and over the land is $\sim$ 11 W m$^{-2}$. We define a BSISO event to be propagating if the lowest value of box-averaged 25-70 day filtered OLR anomaly is below -17.2 W m$^{-2}$ in the EIO box, and after attaining the lowest value in the EIO, it attains its lowest value over the land box within next 20 days and it is below -13 W m$^{-2}$, which is slightly lower than one standard deviation of this box. These criteria allow us to isolate cases that propagate northward with a substantially strong convective signal, at the same time it doesn't significantly affect the number of cases selected. Note that, Day 0 is defined to be when the box-averaged OLR anomaly attains a minima in the EIO box, and it has to fall within 20 May to 10 Oct.  
The time interval between the EIO box minima to land box minima for the propagating cases is between 10-20 days. 
Using these criteria, we obtain a total of 35 propagating cases from 30 years. 

\noindent Similarly, to isolate non-propagating events, we use the same criteria over the EIO box, 
but constrain the lowest value in the land box within subsequent 20 days to be above -6.5 W m$^{-2}$ , i.e., half of the cut-off value taken for the propagating cases, and also slightly higher than half of the standard deviation of the land box. Thus, we are able to isolate cases that started with almost an equal strength in the EIO but didn't propagate over the SA region with substantial convective signal. This resulted in 15 such cases from 30 years of data. Though the number of non-propagating cases are quite small, slightly loosening the cut-off to -7.5 W m$^{-2}$ picks up 19 cases, and this criteria also doesn't change any basic results, but we prefer to use criteria that are quite strict for isolating propagating and non-propagating cases, because they identify very distinct cases, which is helpful for comparing their propagation characteristics. We confirm that the results are insensitive to slight changes in the box size, location, and threshold values.

\noindent After isolating the propagating and non-propagating cases, we construct composites. To understand the moist processes associated with the movement of convection, and the evolution of circulation, we perform a moisture and vorticity budget analysis, respectively. Detailed descriptions of budgets are given in their respective sections. 
For all the constructed composites, we have performed a student's-t test, and we have used 90\% confidence level as numbers of events are not very large. OLR and winds are shown only when they are significantly different from zero at 90\% confidence level, while for the budget related calculations, we have plotted the full signal, but we confirm that all the signals at almost all the pixels in the location of interest are statistically significant at 90\% confidence level. 

\section{Horizontal Structure of the two types of BSISO events}

\begin{figure}
    \centering
    \includegraphics[width=\textwidth]{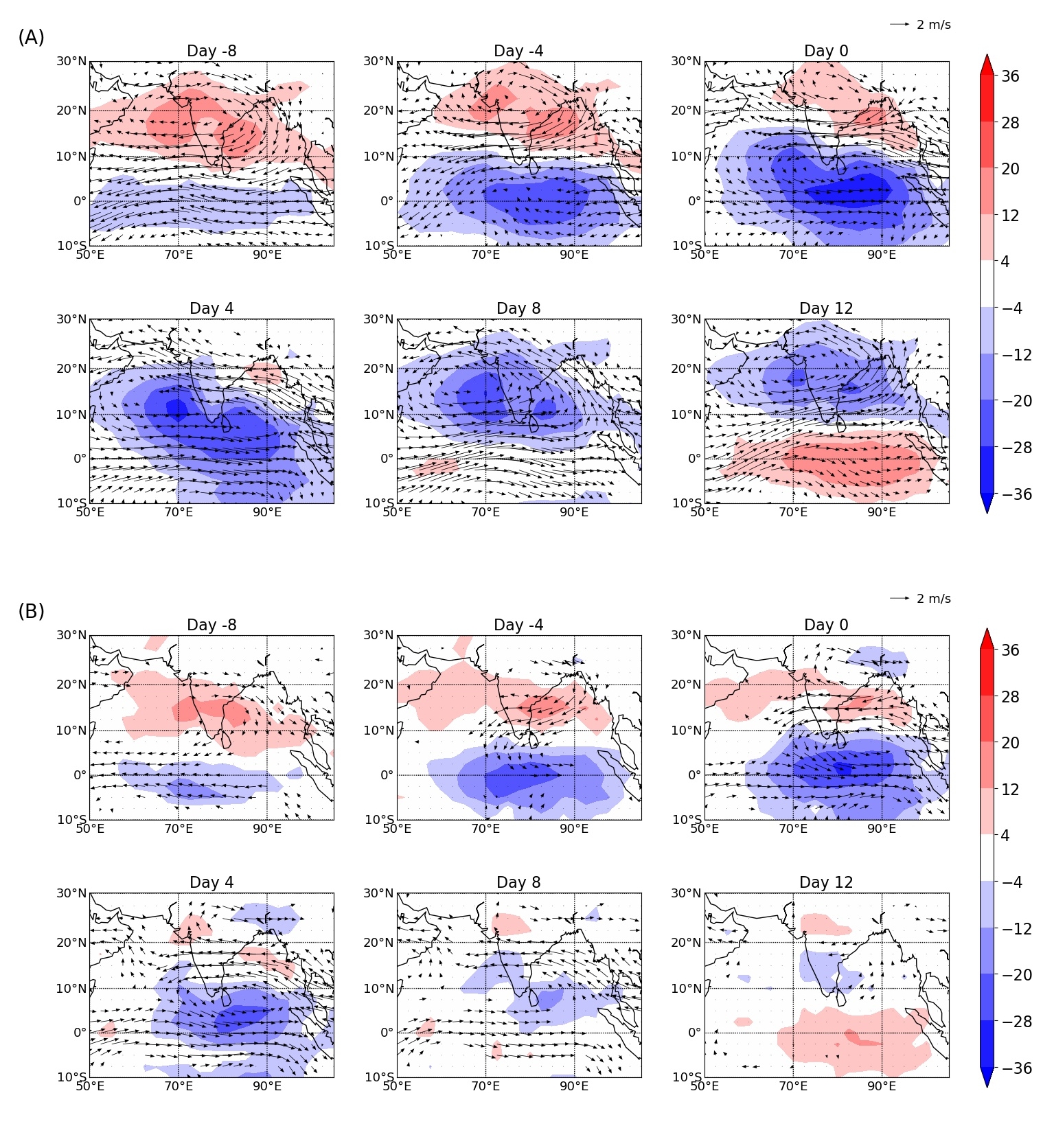}
    \caption{A. Composite of $25-70$ day filtered OLR (W m$^{-2}$; shading) and 700 hPa wind anomalies (quivers) from Day -8 to Day 12 for the propagating cases. B. Same as (A) but for non-propagating events. OLR and wind vectors shown are statistically significant at 90\% confidence level.}
    \label{fig1}
\end{figure}

\noindent In this section, we present the horizontal composite characteristics of the propagating and non-propagating BSISO events. We begin with the propagating composite (Figure \ref{fig1}A); specifically, we show OLR and 700 hPa (lower free-troposphere) horizontal wind anomalies with an interval of four days. On Day -8 (Figure \ref{fig1}A), we see positive OLR anomalies that engulf the entire Arabian Sea (AS), the Bay of Bengal (BOB), and a significant part of the South Asian landmass, while over the EIO, there is a hint of enhanced convection (weak negative OLR anomaly). Associated with the positive OLR anomalies, we see an anticyclonic gyre (slightly tilted from NW-SE), and easterlies over the EIO and the AS. These wind anomalies can be understood as a response to convective heating anomalies modified by monsoon background. Details about the circulation anomalies and their movement will be discussed later as this will turn out to be an important question as we proceed in the paper.

\noindent The enhanced convection gets stronger by Day -4 and it crosses 10N into the AS, simultaneously, the existing positive OLR anomalies in the AS recede. Interestingly, over the BOB, the positive OLR anomalies (suppressed convection) don't recede as much. At this time, we see hints of a new cyclonic Rossby-type gyre forming over the EIO, slightly north of the enhanced convection, while the anticyclonic circulation continues over the land region, thus, between 5-20N, to the north of the nascent convective anomalies,  we observe strong anomalous easterlies. A Kelvin wave response can also be seen further east (not shown), but as we are focusing on the South Asian Monsoon Region, we limit our focus to the Rossby part of the response. On Day 0, the newly formed cyclonic Rossby gyre gets more pronounced in response to the strengthened convective anomaly in the EIO; with strong easterlies north of the convection and westerlies into the convection near the equator. Further, a weaker Rossby-type circulation is also visible south of the equator. This can be understood as a modified Gill-type response under monsoon background winds to enhanced convective heating anomalies (slightly asymmetric, stronger in the north) at the equator. Interestingly, similar to Day -4, on Day 0 northward march of convection continues into the AS, but it doesn't enter the BOB. Thus, the convection that started with a zonally oriented structure in the EIO on Day -8, progresses deep into the AS by Day 0, but not in the BOB, and a clear NW-SE tilted structure gets established. The area of suppressed convection to the north shrinks further, essentially vanishes from the AS and weakens over land, though it is still prominent over the BOB.   

\noindent On Day 4 (Figure \ref{fig1}A), positive OLR anomalies over land vanish, convection enters deep into the northern AS and also moves slightly north over the BOB; moreover, the NW-SE tilt of the convection band becomes pronounced. The most interesting feature of this day is the circulation pattern --- the cyclonic Rossby gyre that formed over the EIO in response to the enhanced convection gets abruptly tilted from NW to SE and moves north, thus the wind anomalies over land become south-easterly. In other words, the cyclonic vortex quickly ``jumps" north into the AS between Day 0 and Day 4, but in contrast, it moves slowly in the BOB sector. On Day 8, anomalous convection enters deep over the land region, and engulfs peninsular, central, and north-west India. The tilted structure of the cyclonic vortex becomes more prominent and moves further north, accompanying which are pronounced south-easterlies over land.
Over the EIO, from Day 0 onwards, we see westerlies associated with the southern flank of the cyclonic vortex, and the convection dies down by Day 8. Essentially, the entire convective belt moves north from the EIO and by Day 12, anomalous convection covers most of the South Asian landmass. In the EIO, the next cycle of the BSISO starts as positive OLR anomalies appear over the region. Indeed, this leads to the period of approximately 40 days for this mode. 

\noindent Shifting our focus to the composite of the non-propagating cases (Figure \ref{fig1}B) --- on Day -8, we see an area of suppressed convection to the north of 10N, engulfing the AS, peninsular India and part of the BOB, but the strength and extent of the suppressed convection as well as associated wind anomalies are much weaker compared to its propagating counterpart. In the EIO, we see a hint of enhanced convection with a narrow band of negative OLR anomalies. By Day -4, the area of enhanced convection increases and becomes stronger, and a small but well-formed Rossby gyre comes into being to the north of this equatorial convection as a response to this convective heating. Interestingly, the area of suppressed convection north of the nascent enhanced convection doesn't weaken much. On Day 0, the equatorial convection strengthens, and the Rossby gyre is more prominent with westerlies over the EIO and easterlies between 10-20N, and a weaker gyre can also be seen south of the Equator. Again, such response can be understood as a modified Gill-type response.  The difference with the Rossby gyre of the propagating cases is in their extent; specifically, the easterlies associated with the gyre of the propagating composite (Figure \ref{fig1}A) extends west of 70E, while for the non-propagating cases, they are mostly confined to the east of 70E. On Day 4, enhanced convection in the EIO remains almost stationary (though slightly weaker), along with the Rossby gyre. While the area of suppressed convection gets weaker, particularly in the northern AS, strong convective signal barely moves into the AS which is in stark contrast to the propagating cases. Moreover, unlike the propagating cases, the Rossby gyre doesn't get tilted towards the north-west.  Subsequently, from Day 8 to Day 12 (Figure \ref{fig1}B), the convective signal weakens and finally disappears, and a new area of suppressed convection appears over the EIO. Thus, in spite of having a strong convective signal in the EIO, the BSISO doesn't propagate northward but stalls and dies out.

\section{Moisture budget}

 \noindent The tropical atmosphere is known for weak spatial and temporal variability in temperature and large variability in free-tropospheric moisture \citep{charney1963note,sobel2000modeling,sobel2001weak}. Further, tropical convection is known to be tightly coupled with column-integrated moisture on various timescales \citep{bretherton2004relationships,rushley2018reexamining,wolding2020interactions}; in particular, organized convection is highly associated with low-to-middle free-tropospheric moisture \citep{sherwood1999convective,holloway2009moisture,takayabu2006diagnostic}. Many studies regarding the MJO/BSISO have shown the coherence between anomalous convection and column-integrated moisture (specific humidity)/MSE/ME anomalies \citep{kiranmayi2011intraseasonal,adames2015three,kim2014propagating,jiang2018unified,gao2019diagnosing,wang2020diagnosing}, or sometimes moisture anomalies slightly lead the anomalous convection \citep{yasunaga2012differences,kim2017does}. Following such observations, the moisture mode framework has emerged as a promising avenue to understand large-scale moist tropical ISOs, which at the broadest level means that these modes of variability are primarily dictated by evolution of moisture anomalies and that these modes would not exist in any model that does not contain a prognostic equation for moisture \citep{sobel2014moist,kim2014propagating,jiang2020fifty}. In contrast, horizontal or temporal variations of temperature may be ignored, which is known as the weak temperature gradient (WTG) approximation \citep{sobel2001weak,sobel2000modeling,kim2014propagating}. Thus, a moisture budget has been used to understand the processes involved in BSISO \citep{adames2016seasonality,chen2021diversity} and MJO \citep{nasuno2015moistening,adames2015three,adames2016seasonality,kang2021role} dynamics. Moist static energy (MSE)/Moist entropy (ME) budgets have also been employed \citep{kiranmayi2011intraseasonal,kim2014propagating,pillai2016moisture,jiang2018unified,wang2020diagnosing}, as these include additional energy fluxes that may affect convection and MSE/ME are nearly conserved variables. These studies concluded that MSE/ME anomalies are in fact controlled by moisture anomalies, consistent with the WTG approximation. 
 Inspired by this framework, we use a moisture budget analysis to understand the essential reasons behind the northward propagation of BSISO.

\begin{figure}
    \centering
    \includegraphics[width=\textwidth]{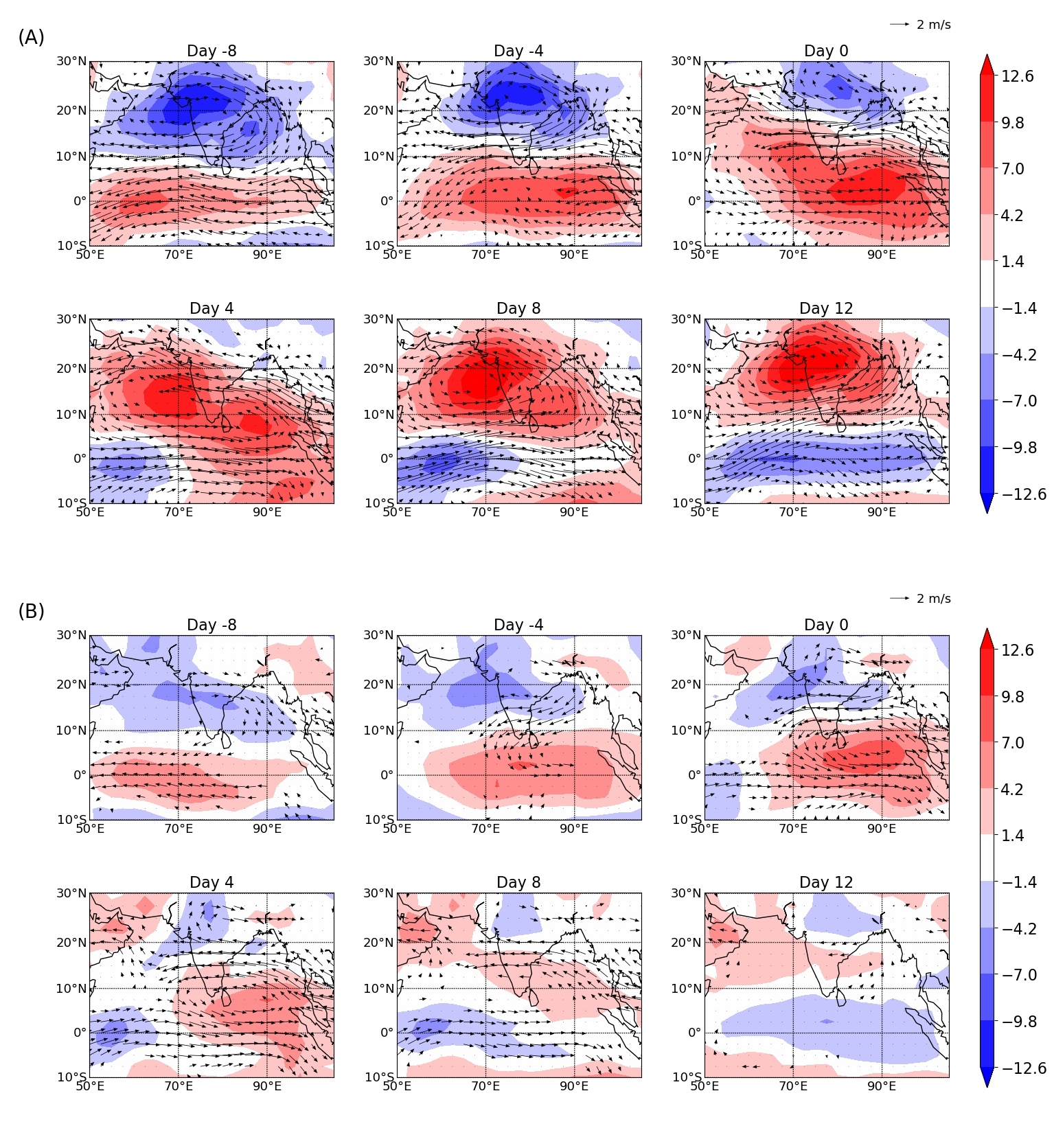}
    \caption{A. Composite of 25-70 day filtered column-integrated specific humidity (scaled by the latent heat of vaporization) ($10^{6}$J m$^{-2}$; shading) and 700 hPa wind anomalies (quivers) from Day -8 to Day 12 for the propagating cases. B. Same as (A) but for non-propagating events. Wind vectors shown are statistically significant at 90\% confidence level.}
    \label{fig2}
\end{figure}

\noindent Before jumping into the moisture budget analysis, we show 25-70 day filtered column-integrated (1000 to 200 hPa) specific humidity anomalies for propagating (Figure \ref{fig2}A) and non-propagating (Figure \ref{fig2}B) BSISO cases. Comparing with Figure \ref{fig1} and \ref{fig2}, we clearly see that BSISO-related specific humidity and OLR (convection) anomalies are generally collocated, though, to the north of 20N, including over the land, moisture anomalies slightly lead the OLR anomalies (see also \cite{jiang2018unified,gao2019diagnosing,wang2020diagnosing,masiwal2023contrasting}), which possibly indicates the need for some pre-moistening before the onset of deep convection in those regions. Proceeding to the moisture budget, the relevant equation reads, 
\begin{equation}
[\frac{\partial q'}{\partial t}] = - [(\textbf{V}.\nabla_{h} q)]'-[(\omega\frac{\partial q}{\partial p})]'-P'+E'+R, 
\label{e1}
\end{equation}
where $q$ is the specific humidity, $\textbf{V}=u\textbf{i}+v\textbf{j}$  is the horizontal wind,  ${\nabla_{h}=\textbf{i}(\frac{\partial }{\partial x})+\textbf{j}(\frac{\partial }{\partial y}})$ is the horizontal gradient operator, $P$ is  precipitation, $E$ is  evaporation, and $\omega$ is the vertical velocity in pressure co-ordinates. Here, prime denotes a 25-70 day(BSISO) anomaly. $R$ is the residual in the budget. The square bracket represents mass-weighted vertical integrals, calculated from 1000 to 200 hPa. The last three terms of the R.H.S are usually bundled together as, 
\begin{equation}
-[Q_2]'/L=-P'+E'+R,\label{e2}
\end{equation}
which is often called the column-integrated ``apparent moisture sink" \citep{adames2015three}. The last four terms in Equation \ref{e1} are together called ``column-processes" \citep{kang2020role}. Further, 
\begin{equation}
C' = -[(\omega\frac{\partial q}{\partial p})]'-P'+E'+R=-[(\omega\frac{\partial q}{\partial p})]'-[Q_2]'/L.
\label{e3}
\end{equation}
Hence, this term can be calculated directly by subtracting horizontal advection from moisture tendency. As precipitation and evaporation are not defined at pressure levels, the moisture budget equation at a single pressure level is often written as, \begin{equation}\frac{\partial q'}{\partial t} = - (\textbf{V}.\nabla_{h} q)'-(\omega\frac{\partial q}{\partial p})'-Q'_2/L, 
\label{e4} \end{equation} which we have used for our vertical structure investigation. Moreover, the vertical moisture advection term can be broken down into,
\begin{equation}-(\omega\frac{\partial q}{\partial p})'=-\frac{\partial(\omega q)'}{\partial p}-  (q\nabla_.{\textbf{V}})', 
\label{e5} \end{equation} 
where the two terms in the RHS are the vertical moisture flux convergence and horizontal moisture convergence, respectively. Note that, as many previous studies regarding MJO/BSISO used MSE/ME budgets, we multiply all the budget terms with $L$, the latent heat of vaporization as this will prove useful in comparing our results with the literature \citep{kiranmayi2011intraseasonal}.  

\noindent As seen in Figure \ref{fig2}A, early in the BSISO cycle, i.e., from Day -8 to Day -4, there is the first hint of northward movement of the moisture anomaly (as well as convection; Figure \ref{fig1}A) from the EIO to the southern AS along with a weakening of suppressed convection in the northern AS. This northward movement of convection over the AS continues, and on Day 0, we see that the positive moisture anomaly has almost engulfed the AS up to 17.5 N. Strikingly, from Day -8 to Day 0, there is very little northward movement of moisture anomalies over the BOB, resulting in a characteristic NW-SE tilted convection band associated with the developing BSISO. 

\begin{figure}
    \centering
    \includegraphics[width=\textwidth]{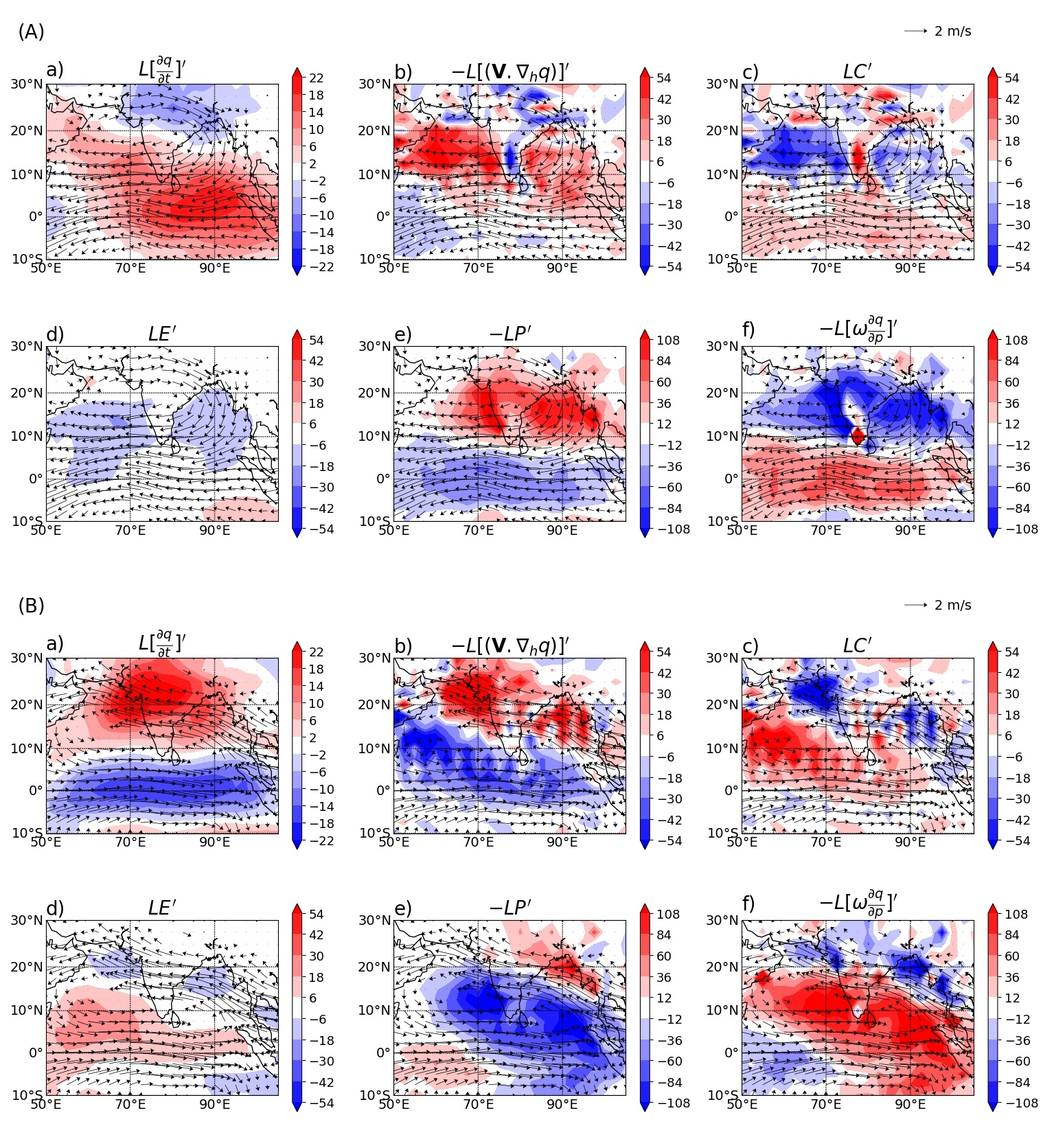}
    \caption {A. Contours of the composite mean of 25-70 day anomalous  terms of the column-integrated moisture budget as in Equation \ref{e1} (scaled by the latent heat of vaporization) and their combinations (for column-processes) for propagating cases on Day -8, including, (a) Moisture tendency, (b) Horizontal advection,(c) Column-processes, (d) Evaporation, (e) Precipitation, (f) Vertical advection. Units of terms are W m$^{-2}$. The 700 hPa wind anomalies are overlaid for reference. Wind vectors shown are statistically significant at 90\% confidence level. B. Same as (A), but for Day 4.}
    \label{fig3}
\end{figure}

\noindent To understand the reason behind the preferential northward propagation over the AS at this stage, we study the moisture budget on Day -8 as shown in Figure \ref{fig3}A. As expected, the moisture tendency (Figure \ref{fig3}(A)a) has entered into the AS, and the moisture budget terms indicate that horizontal advection (Figure \ref{fig3}(A)b) is the main contributor to the moisture tendency in the AS sector, though it is not the dominant term in the budget.  Precipitation (Figure \ref{fig3}(A)e) and vertical advection (Figure \ref{fig3}(A)f) are the dominant terms, but they cancel each other to a large extent \citep{adames2015three}. The large precipitation and vertical advection anomaly signals are due to the previous cycle of suppressed BSISO convection north of the newly born convective anomaly over the EIO. Evaporation (Figure \ref{fig3}(A)d), though small in magnitude, opposes the moistening, because easterly wind anomalies associated with the BSISO act against the climatological south-westerly monsoonal mean winds and slow down the overall flow in the AS as well as BOB. This has been noted previously \citep{sengupta2001coherent,gao2019diagnosing,wang2020diagnosing}, and indicates that the WISHE mechanism is not applicable to the northward propagation of the BSISO. 

\noindent Amongst ``column-process" terms (Figure \ref{fig3}(A)c), vertical advection dominates and results in a net negative value over AS and BOB, with a small positive contribution over the EIO. Both ``column-process" and horizontal advection moisten the EIO, while the ``column-process" dries the AS and BOB in this stage. In BOB, the magnitude of horizontal advection is close to the opposing ``column-process", and in all, we see a drying tendency in the north and weak moistening in the southern BOB. But over AS, horizontal advection is considerably larger, and it wins over the drying ``column-process", and the net result is a moistening of the atmosphere over the AS. This is the primary cause for the preferential movement of moist convection over the AS at this stage in its development. These findings, at this stage of the BSISO, are consistent with previous ME/moisture budget analyses \citep{jiang2018unified,chen2021diversity}. In peninsular India, we see a slightly different moistening process. Unlike AS and BOB, the horizontal advection on Day -8 and Day -4 dries the region. Yet, we find overall moistening as reflected in the tendency term, indeed, this is due to ``column-process", more specifically, due to vertical advection. A similar pattern was also identified by \cite{jiang2018unified} who speculated that this is probably due to topographic influence. As seen in Figure S1, Day -4 tells a very similar story about the development of the moisture anomalies.

\noindent Next, we focus on the moistening over the South Asian landmass, particularly around and beyond 20N. As seen in Figure \ref{fig2}A,  the most striking moisture build-up happens between Day 4 and Day 12. To understand this moistening we examine the moisture budget on Day 4 in Figure \ref{fig3}B. Clearly, we see a very strong moisture tendency covering almost all the South Asian landmass (Figure \ref{fig3}(B)a) above 15N. Similar to Day -8, horizontal advection  (Figure \ref{fig3}(B)b) is the main contributor to the positive moisture tendency. We see that the ``column-process" ( Figure \ref{fig3}(B)c; again, mainly due to vertical advection, and to some small extent, evaporation) induces drying over the northern AS and the desert region of India and Pakistan, but that is much weaker than the horizontal advection. The BOB also shows a large positive moisture tendency, and it is also due to horizontal moisture advection, which is now considerably stronger than the ``column-process''. The story is very similar for Day 8 (Figure S2), the land above 20N gets further moistened by horizontal advection, though ``column-process'' due to vertical advection has a minor contribution too. This result is in line with the observation that over the Indian subcontinent, horizontal advection of moisture leads the abrupt transition from shallow to deep convective clouds associated with BSISO by a few days \citep{wang2015cloud}. 


\noindent For the non-propagating cases, the composite in Figure \ref{fig2}B shows that from Day -8 to Day 0, the anomalous  moisture distribution remains almost unchanged  over the AS, while there is a strengthening of the positive anomaly over the EIO. Even on Day 4, when the moisture anomaly at the EIO starts weakening, though we see a sign of moistening over the AS (weakening of negative anomaly), but, the positive moisture anomaly fails to penetrate into the AS. Thus, in stark contrast to the propagating cases, even when the EIO convection is the strongest, a positive moisture anomaly doesn't penetrate into the AS. The situation remains almost unchanged on Day 8 and Day 12 (Figure \ref{fig2}B), though now we see the sign of a very weak positive moisture anomaly over the AS, but it is much weaker when compared to the propagating cases, and fails to trigger deep convection. Further, over the EIO, positive moisture anomalies almost vanish by Day 8. This is also evident in the OLR signal in Figure \ref{fig1}B. So, the first noticeable major contrast from the moisture (OLR) anomaly perspective between the propagating and non-propagating cases is, when the convection in the EIO is strongest, moisture (OLR) anomalies enter into the AS for the propagating cases, while for the non-propagating cases, they fail to penetrate into the AS, and the convection eventually dies down. Thus, failure to propagate into the AS also predicates the failure to propagate over land in the next stage. In a way, one can state that the success/failure of strong moisture anomalies to penetrate into the AS seals the fate of the newborn equatorial BSISO convection, i.e., whether it can propagate northward over SA or not. As the convection signal in EIO is strongest on Day 0, and the first well-organized and strong circulation appears on that day, we focus on the moisture budget of day Day 0 for non-propagating BSISO composites to find out why the convection fails to propagate in the AS.

\begin{figure}[h]
    \centering
    \includegraphics[width=\textwidth]{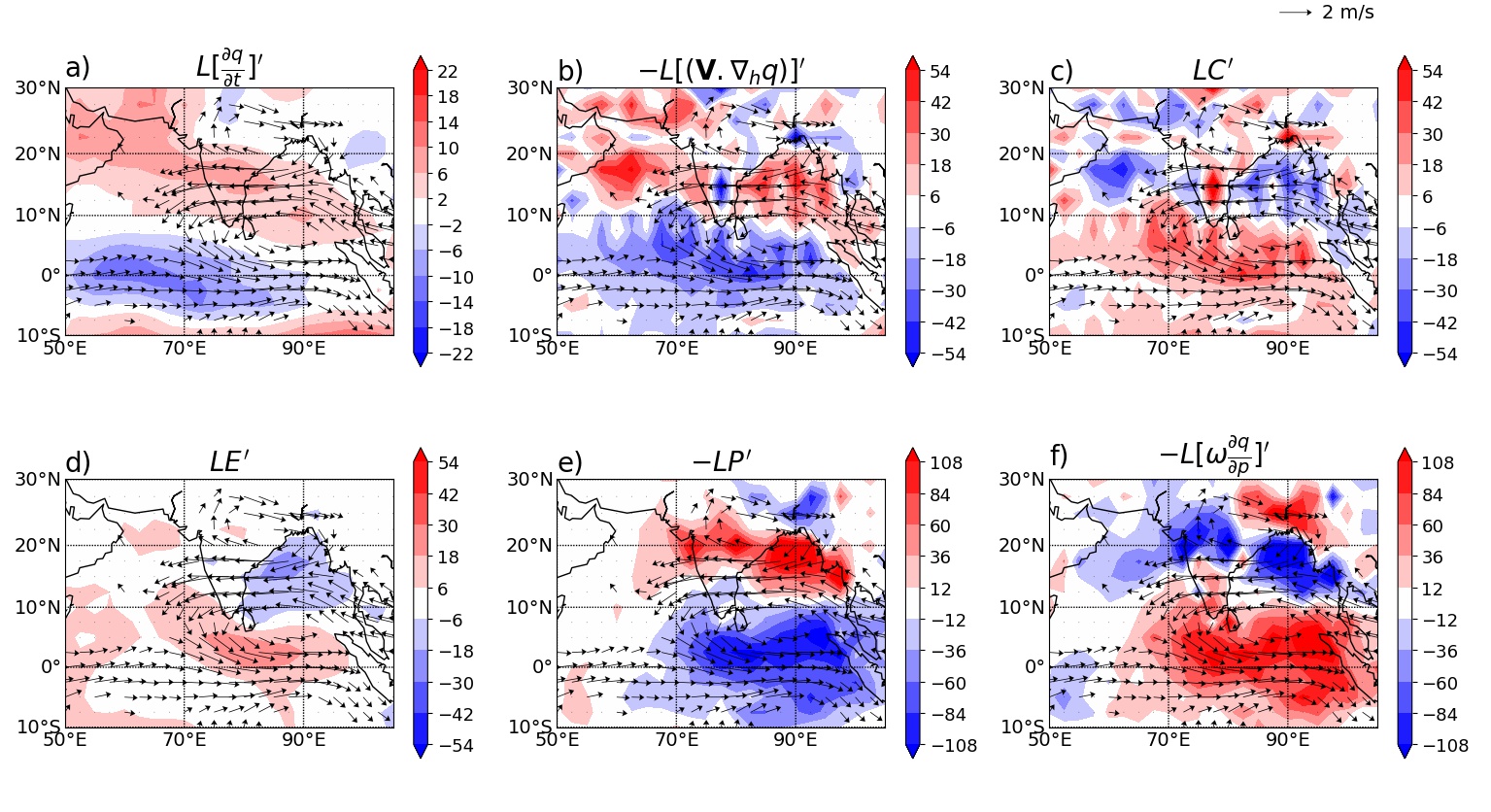}
    \caption{Same as Figure \ref{fig3}, but for Day 0 of the non-propagating cases.} 
    \label{fig4}
\end{figure}

\noindent The moisture budget on Day 0 for the non-propagating cases is shown in Figure \ref{fig4}. As in the propagating BSISO cases, the tendency (Figure \ref{fig4}a) is dominated by the horizontal advection  (Figure \ref{fig4}b) and the ``column-process" (Figure \ref{fig4}c) (mostly by vertical advection) acts against it to reduce the amount of moistening. But here, the tendency is much weaker than the propagating cases, particularly over the southern AS. This weak tendency is caused by the weak horizontal advection term (in some areas near 10S, it is even slightly negative), particularly in the southern AS. One can compare this horizontal advection with the horizontal advection of Day -8 (Figure \ref{fig3}(A)b) of propagating composite to appreciate the stark contrast. This weak moistening is the reason behind the inability of the convection to move into the AS. Interestingly, the moistening is slightly stronger in the northern AS, which is also reflected in the fact that from Day 0 to Day 4, the negative moisture anomaly in the northern AS weakens, but the positive anomaly fails to penetrate into the southern AS. Over peninsular India, advection is negative, and ``column-process" (dominated by vertical advection) is positive, and we see a net weak positive tendency. Finally, over the EIO, horizontal advection induces drying, and that eventually kills the convection in that region.

\subsection{Vertical structure}

\begin{figure}
    \centering
    \includegraphics[width=\textwidth]{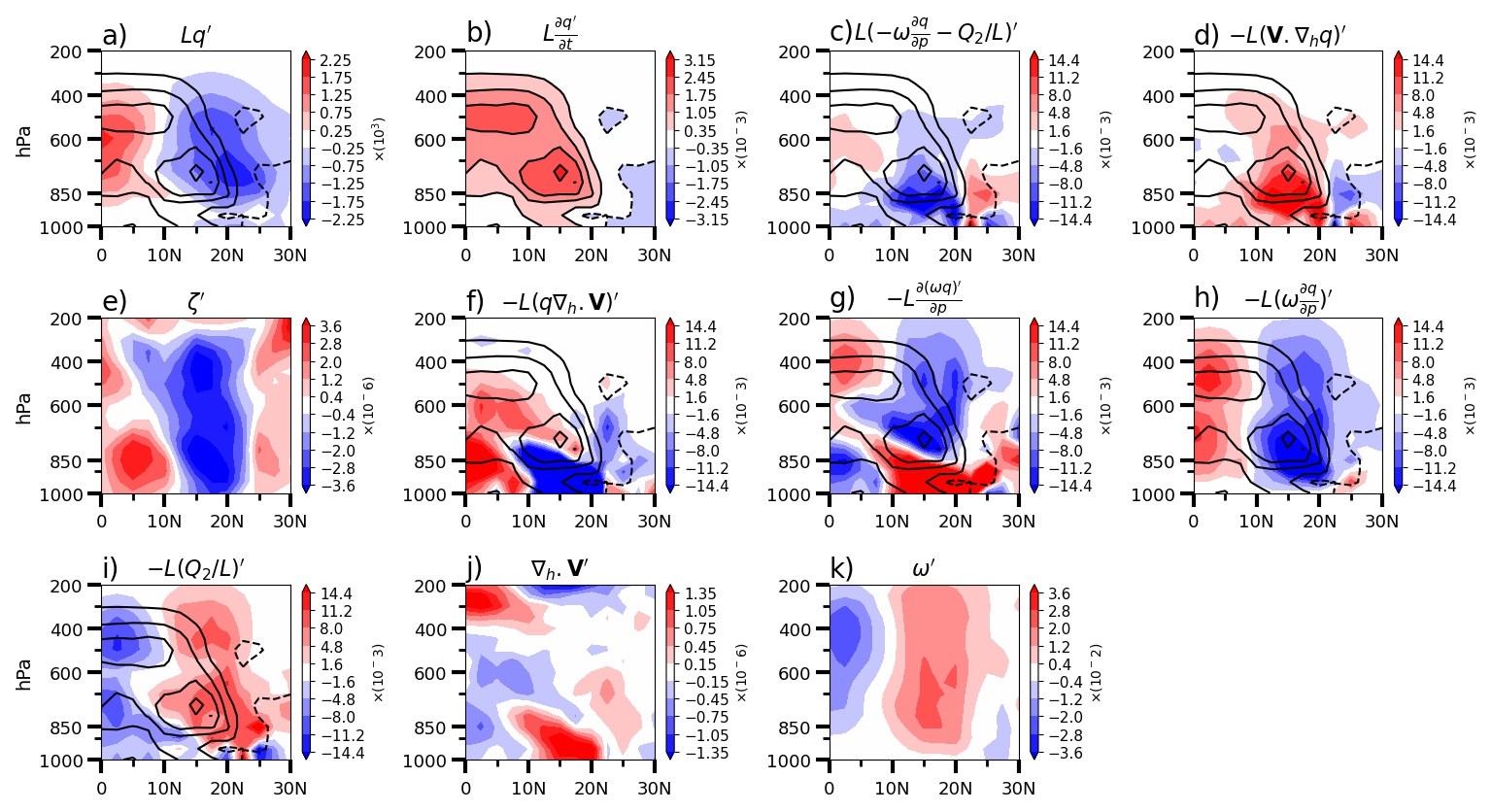}
    \caption{Pressure-latitude profiles of 25-70 anomalous moisture budget terms in Equation \ref{e4} (scaled by latent heat of vaporization), components of anomalous vertical advection decomposition as shown in Equation \ref{e5} (scaled by the latent heat of vaporization), and anomalies of a few other important variables averaged over target region of Arabian Sea  (60$^\circ$-72.5$^\circ$E) on Day -8 for propagating cases. The terms are namely (a) Specific humidity, (b) Moisture tendency, (c) The sum of vertical moisture advection and apparent moisture sink,  (d) Horizontal moisture advection, (e) Vorticity, (f) Horizontal moisture convergence, (g) Vertical moisture flux convergence, (h) Vertical moisture advection, (i) Apparent moisture sink, (j) Divergence, and (k) Vertical velocity in pressure coordinates. Moisture tendency contours are overlaid on moisture budget terms. Units for specific humidity (moisture) is J kg$^{-1}$, vorticity and divergence is s$^{-1}$, vertical velocity is Pa s$^{-1}$, and for all other terms is W kg$^{-1}$.} 
    \label{fig5}
\end{figure}

\noindent Having understood the column-integrated moisture budget, we now examine the vertical structure of the BSISO. From the discussion above, the critical difference between the propagating and non-propagating cases stems from the ability/inability of the deep convection to penetrate into the AS, thus, here we focus on the AS sector (60-72.5E). Figure \ref{fig5} shows the vertical structure of the moisture budget and other important terms on Day -8 for propagating BSISO events, when moistening starts over the AS. Note that vertical advection is decomposed into two parts including horizontal moisture convergence, shown in Equation (\ref{e5}), as we want to categorically focus on BLMC, which is thought to be critical for northward propagation \citep{jiang2004structures,bellon2008instability}. 
Moreover, the vertical structure enables us to examine whether the BSISO has a pronounced tilt similar to the MJO \citep{hsu2012role,adames2015three,jiang2015vertical}, as implied by the BLMC theories of the BSISO.

\noindent As seen in Figure \ref{fig5}, on Day -8, a positive moisture anomaly (Figure \ref{fig5}a) is located over the EIO, south of 10N, and a negative moisture anomaly is present to the north over the AS. Both anomalies reach up to 400 hPa as seen in Figure \ref{fig5}. One should note that the strongest moisture anomaly is not in the boundary layer, but just above it in the free-troposphere. In fact, unlike the MJO \citep{hsu2012role,adames2015three,jiang2015vertical}, BSISO-related moisture anomaly as well as tendency is very weak in the boundary layer (see also \cite{neena2017model}), this itself indicates BLMC can't be important for moistening. As expected, the positive moisture tendency (Figure \ref{fig5}b) is to the north of the positive moisture anomaly, and interestingly, it is also largest in the lower free-troposphere between 850-600 hPa. The pattern is very similar on Day -4 too (Figure S3), when the moistening becomes much stronger. In Figure \ref{fig5}b, we see another maxima in the tendency term near 500 hPa between 0-10N (which gives the tendency a southward tilt aloft), but we don't focus on it here as it is connected to the strengthening of the nascent convection in the EIO, and our focus is on the propagation mechanism. As the convection enters into the AS by Day -4 (Figure S3), we find the tendency to be mostly upright with only one maxima between 850-600 hPa in the lower free-troposphere. As noted in the previous section, the main contributor to the tendency to the north of the existing positive moisture anomaly is horizontal advection (Figure \ref{fig5}d), while ``column-process" (Figure \ref{fig5}c) mostly acts against it to reduce the moistening. Vertical advection (Figure \ref{fig5}h) dominates the ``column-process" and is stronger than the opposing apparent moisture sink (Figure \ref{fig5}i), thus it is responsible for negative values of ``column-process" in the zone of positive tendency ahead of convection. 
A similar view of the vertical structure has been noted in a modeling study of the BSISO \citep{sooraj2013boreal}. 

\noindent Horizontal advection  and ``column-process" are equally strong in the boundary layer, and cancel each other, while in the free-troposphere, horizontal advection is larger, so the net tendency is positive. Thus, while moistening via boundary layer moisture advection was put forth as a possibility \citep{jiang2004structures}, and received some support \citep{demott2013northward,chou2010mechanisms}, it is clearly not applicable for the BSISO. Decomposing vertical advection, we see positive BLMC does not lead the moisture anomaly. Ahead of the convection, between 10-20N, where the moisture tendency is strongest (positive), large moisture divergence (Figure \ref{fig5}f) inhibits moistening in the boundary layer following the BSISO wind-divergence (Figure \ref{fig5}j). 

\noindent So, from our analysis, it appears that the BLMC mechanism does not play a significant role in moistening the atmosphere ahead of the convection for the BSISO. This is in stark contrast with the MJO where BLMC plays an active role in moistening the boundary layer to the east of active convection \citep{benedict2007observed,hsu2012role,adames2015three}, while horizontal advection moistens lower free-troposphere \citep{kiranmayi2011intraseasonal,kim2014propagating,adames2015three}. Further, in case of MJO, for a given location, boundary layer moistening happens first \citep{benedict2007observed,hsu2012role} via moisture convergence due to a Kelvin wave response which generates shallow convection, following which the moisture anomaly and convection deepen. This phase leading of BLMC with respect to convection as well as moisture anomaly results in a  westward tilt of vertical velocity and moisture with height \citep{benedict2007observed,hsu2012role,adames2015three,jiang2015vertical}. Unlike the MJO, as seen here and also previously noted \citep{abhik2013possible,sarkar2017atmospheric,neena2017model}, a vertical tilt in the vertical velocity and moisture is clearly absent in the BSISO. This is contrary to expectations from BLMC theories \citep{jiang2004structures,bellon2008instability}, as they surmised that BLMC will occur first to the north of the existing convection.

\noindent Moreover, in the BLMC theories \citep{bellon2023selected}, one expects a deep free-tropospheric vorticity structure to lead the convection centre, below which moistening via  BLMC should occur.  This is not the case, as is seen in Figure \ref{fig5}e (or Figure S3). Rather, vorticity is strongest in the boundary layer, and almost out of phase with the area of moistening. Previously, \cite{karmakar2020differences} also questioned the validity of this BLMC view, showing that vertically integrated anomalous vorticity and BSISO precipitation propagates at different speeds over the AS. The evidence presented here, along with the observations presented in the previous section, suggest that BLMC is not a key process in the northward propagation of the BSISO, and BSISO can't be a BLMC moisture mode, rather, it is a traditional moisture mode dominated by horizontal advection in the lower free troposphere.

\subsection{Process of moistening}

\noindent From the moisture budget analyses, we concluded that horizontal advection is the main contributor to moistening that drives the northward propagation. To identify the specific processes responsible for the anomalous horizontal advection 
we decompose it into terms consisting of BSISO-scale and background state wind and moisture components, i.e.,
\begin{equation}(\mathbf{V}.\nabla_{h} q)' \approx (\mathbf{V}'.\nabla_{h} \Bar{q})'+(\mathbf{\Bar{V}}.\nabla_{h} q')'+(\mathbf{V}'.\nabla_{h} q')',
\label{e6}
\end{equation}
where prime means the BSISO-scale perturbation (25-70 day filtered anomaly), and bar refers to the slowly varying seasonal background (mean plus first three harmonics). Of course, there are contributions from other timescales, but they are much smaller than the terms shown, so Equation (\ref{e6}) is a good first-order approximation. In fact, the last term on the R.H.S. is much smaller than the first two terms, and we have observed that these two terms together capture most of the BSISO advection anomaly. Physically, the first term in the R.H.S. is the background moisture advection by the anomalous BSISO wind, and the second term is the anomalous BSISO moisture advection by the background monsoon wind. 

\noindent In the context of moistening in the lower troposphere, we focus on one level (namely 700 hPa). We have chosen this level, as this is where both moisture tendency and moisture advection are large as seen in Figure \ref{fig5}. We also confirm that the column-integrated version of this decomposition paints a very similar picture(not shown). We have chosen the same days we used in our moisture budget analysis to understand the importance of the two processes in different stages.

\begin{figure}
    \centering
    \includegraphics[width=16 cm]{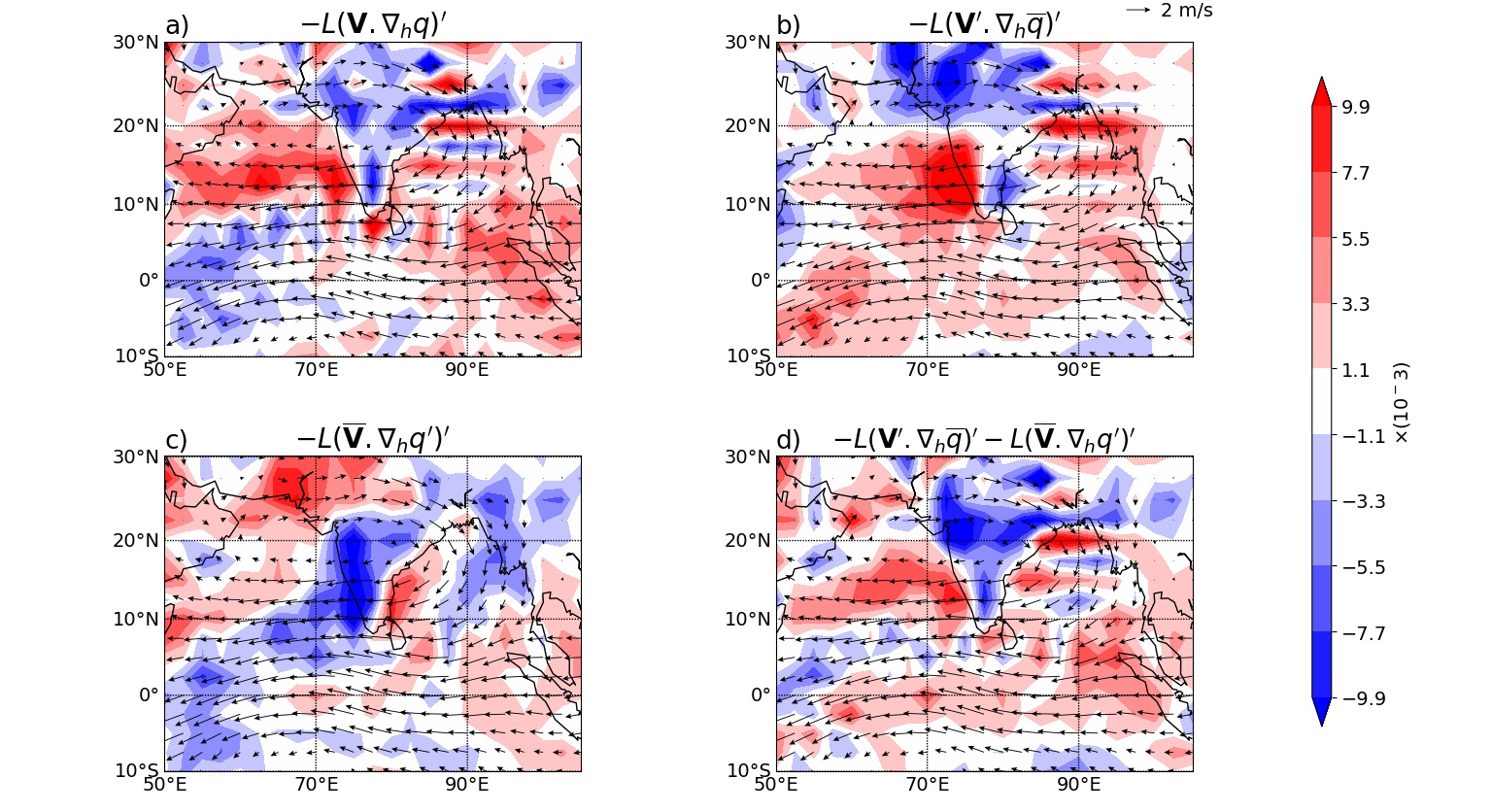}
    \caption{(a) 25-70 day anomalous horizontal moisture advection term and its decomposed primary contributor terms, namely (b) Background moisture advection by BSISO anomalous winds, (c) BSISO anomalous moisture advection by background winds and (d) their combination (all scaled by L) at 700 hPa as shown in Equation 6, for Day -8 of the composite of propagating cases. Units of terms are W kg$^{-1}$. The 700 hPa wind anomalies are overlaid for reference. Wind vectors shown are statistically significant at 90\% confidence level.}   
    \label{fig6}
\end{figure}

\begin{figure}
    \centering
    \includegraphics[width=16cm]{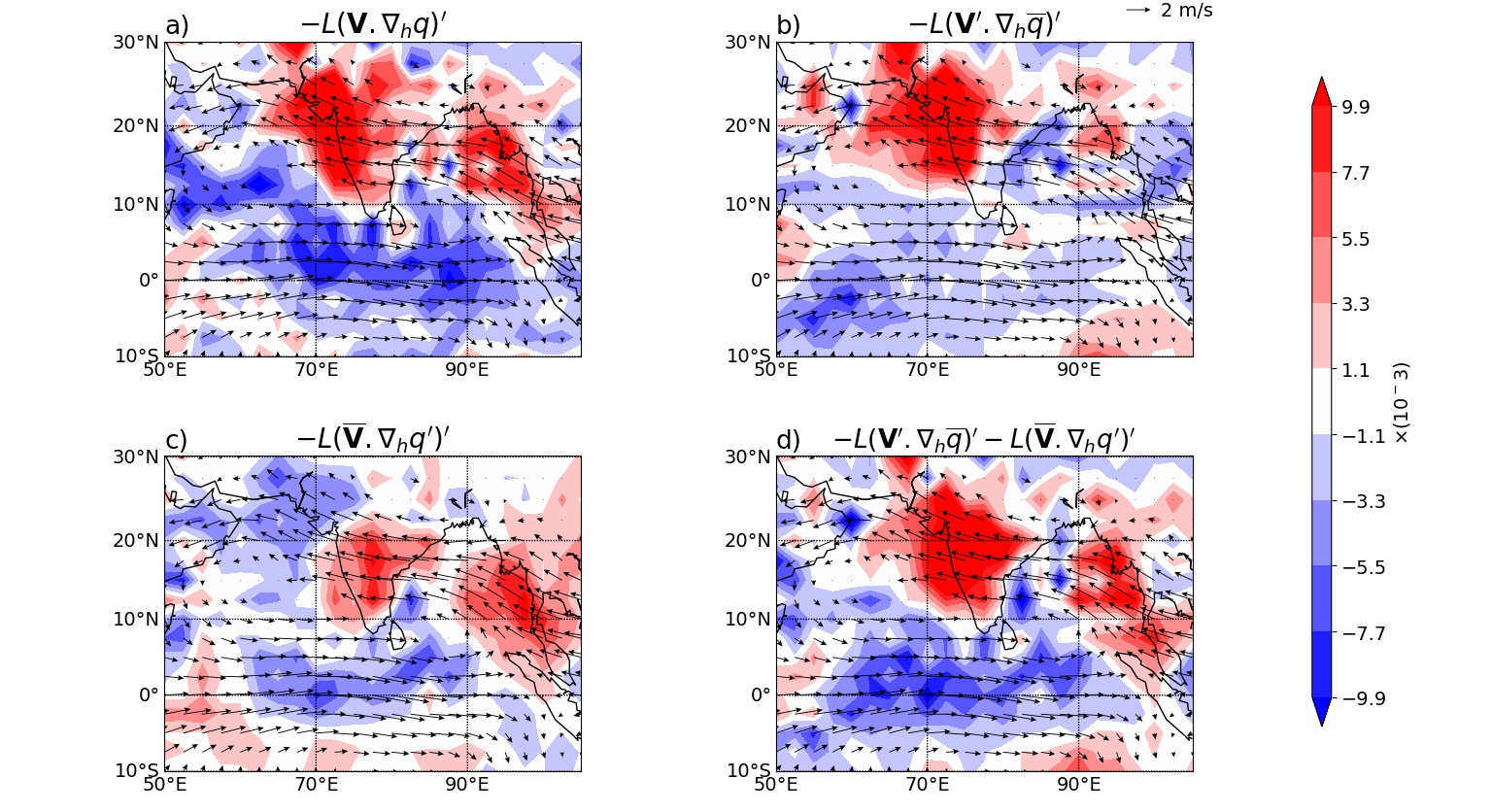}
    \caption{Same as Figure \ref{fig6}, but  for Day $4$.}   
    \label{fig7}
\end{figure}

\noindent From  Figure \ref{fig6} it is clear that on Day -8 of the propagating cases composite, background moisture advection by the BSISO winds (Figure \ref{fig6}b) is the dominant term in the total BSISO horizontal advection (Figure \ref{fig6}a) over the AS sector. The second term (Figure \ref{fig6}c), that is the anomalous BSISO moisture advection by the background monsoon winds has a small contribution to the west of 70E, but it has a negative contribution along India's west coast and peninsular region. Overall, the former is much larger along the west coast, so when added together (Figure \ref{fig6}d), the entire region over the AS has a positive signal. Clearly, Figure \ref{fig6}d captures most of the BSISO-related horizontal advection term (Figure \ref{fig6}a), which proves that the approximation using the first two terms of Equation \ref{e6} holds good. This is true for other days too. In the BOB, both terms have weak contributions, and over the peninsular region advection of the BSISO moisture by background winds dominates, resulting in a net negative anomaly when the terms are added together. On Day -4 (Figure S4), the story is very similar, and entire AS experience even stronger moistening. In the next stage, further north, as seen in Figure \ref{fig7} for Day 4 of the propagating cases, the moistening over land is again dominated by the background moisture advection by the BSISO wind anomalies (Figure \ref{fig7}b). Advection of anomalous BSISO moisture by background monsoon winds tries to dry the northwestern desert region north of 20N (Figure \ref{fig7}c), but it can't overcome the strong moistening by eddy (BSISO) advection of background moisture. The story is different in the peninsular Indian region and BOB, here, background wind advection of BSISO moisture anomalies induces strong moistening (Figure \ref{fig7}c) and dominates the total moisture advection (Figure \ref{fig7}a). Thus, with regard to horizontal advection, both eddy (BSISO) advection of climatological moisture and vice versa are important in different stages and regions of the BSISO. 

\begin{figure}
    \centering
    \includegraphics[width=16 cm]{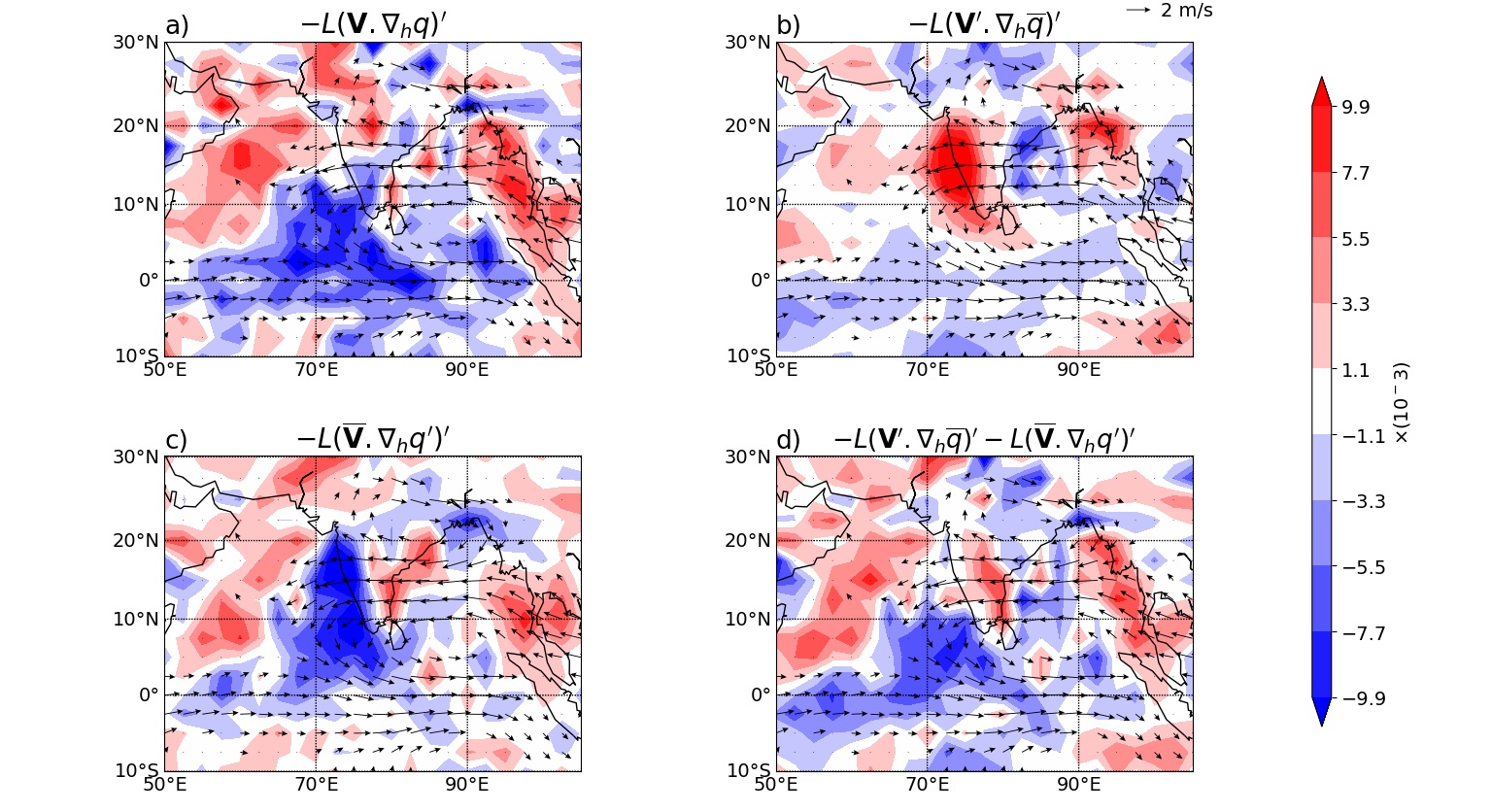}
    \caption{Same as Figure \ref{fig6}, but for Day $0$ of the non-propagating cases.}   
    \label{fig8}
\end{figure}

\noindent Finally, for the non-propagating BSISO cases (Figure \ref{fig8}), the Day 0 maps show that there is net negative moisture advection in the southern AS, particularly from the west coast of India to 60-65E (Figure \ref{fig8}a), which as discussed before as the reason behind the failure of the non-propagating cases to effectively penetrate into the AS. In the north-western AS, to the east of the Arabian peninsula, we see some positive values of horizontal advection. The advection of background moisture by anomalous BSISO winds in the AS is much weaker and limited (close to the coast, east of 70E) compared to the propagating cases, as seen in Figure \ref{fig8}b. Further, the advection of BSISO moisture anomalies by the background wind (Figure \ref{fig8}c) is slightly larger than the propagating cases and essentially offsets any moistening by the advection of background moisture. In fact, this induces a small net negative moisture advection in the southern AS (Figure \ref{fig8}d). So, overall, comparing Figures \ref{fig6} and \ref{fig8}, the difference in horizontal advection of moisture in the southern AS for the non-propagating cases compared to the propagating cases explains the failure of non-propagating cases to penetrate into the AS and is caused by weaker free tropospheric advection of background moisture by weaker and limited BSISO easterlies.

\subsection{Details of Eddy and Mean Flow Advection of Moisture}

\begin{figure}
    \centering
    \includegraphics[width=15 cm]{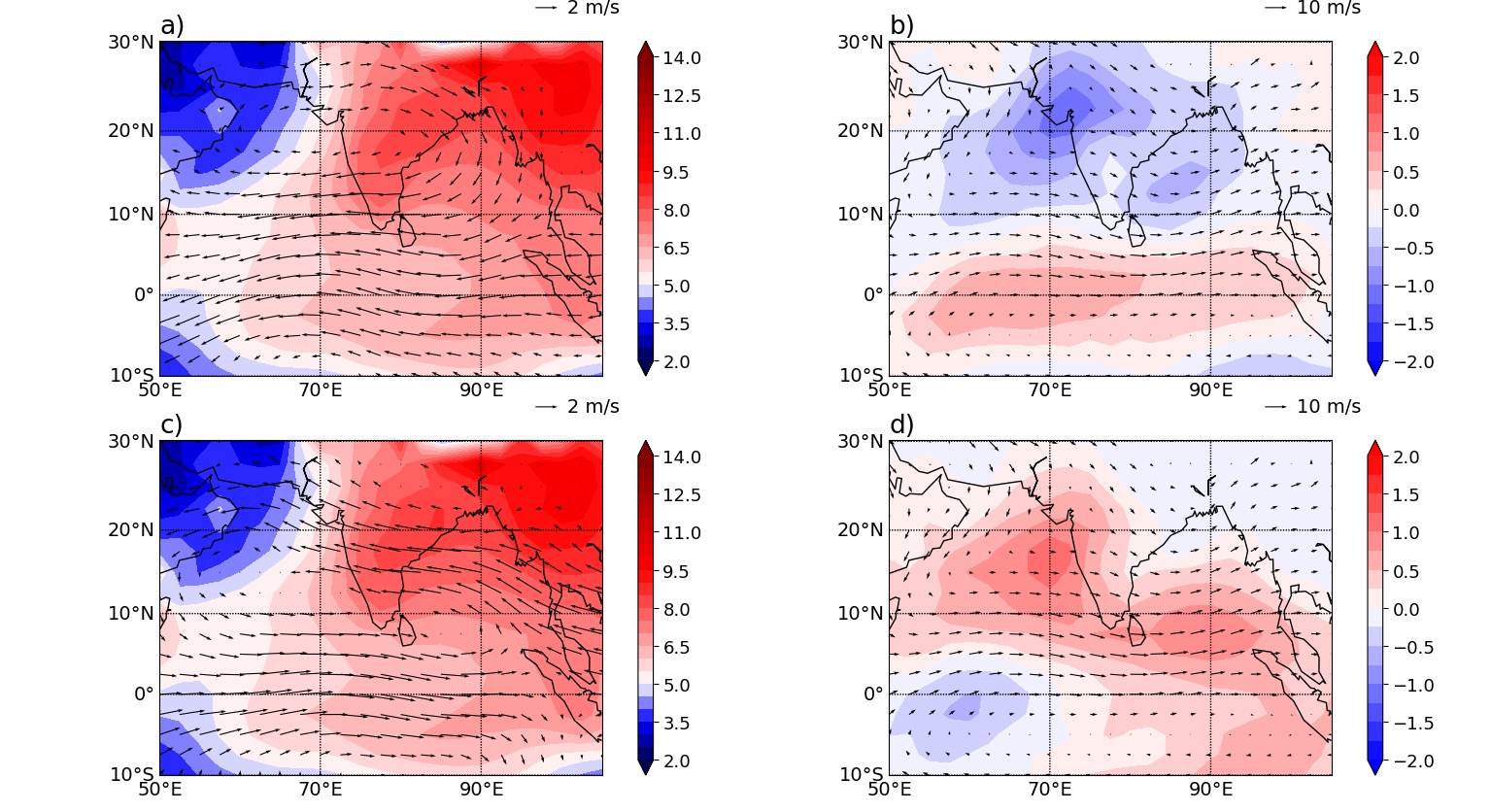}
    \caption{(a) Background specific humidity (g kg$^{-1}$) and 25-70 day filtered wind anomalies, (b) Background winds and 25-70 day filtered moisture anomaly for Day $-8$ of propagating composite at 700 hPa. (c) and (d) are same as (a) and (b) respectively, but for Day $4$.  Wind vectors shown are statistically significant at 90\% confidence level.}  
    \label{fig9}
\end{figure}

\begin{figure}
    \centering
    \includegraphics[width=15 cm]{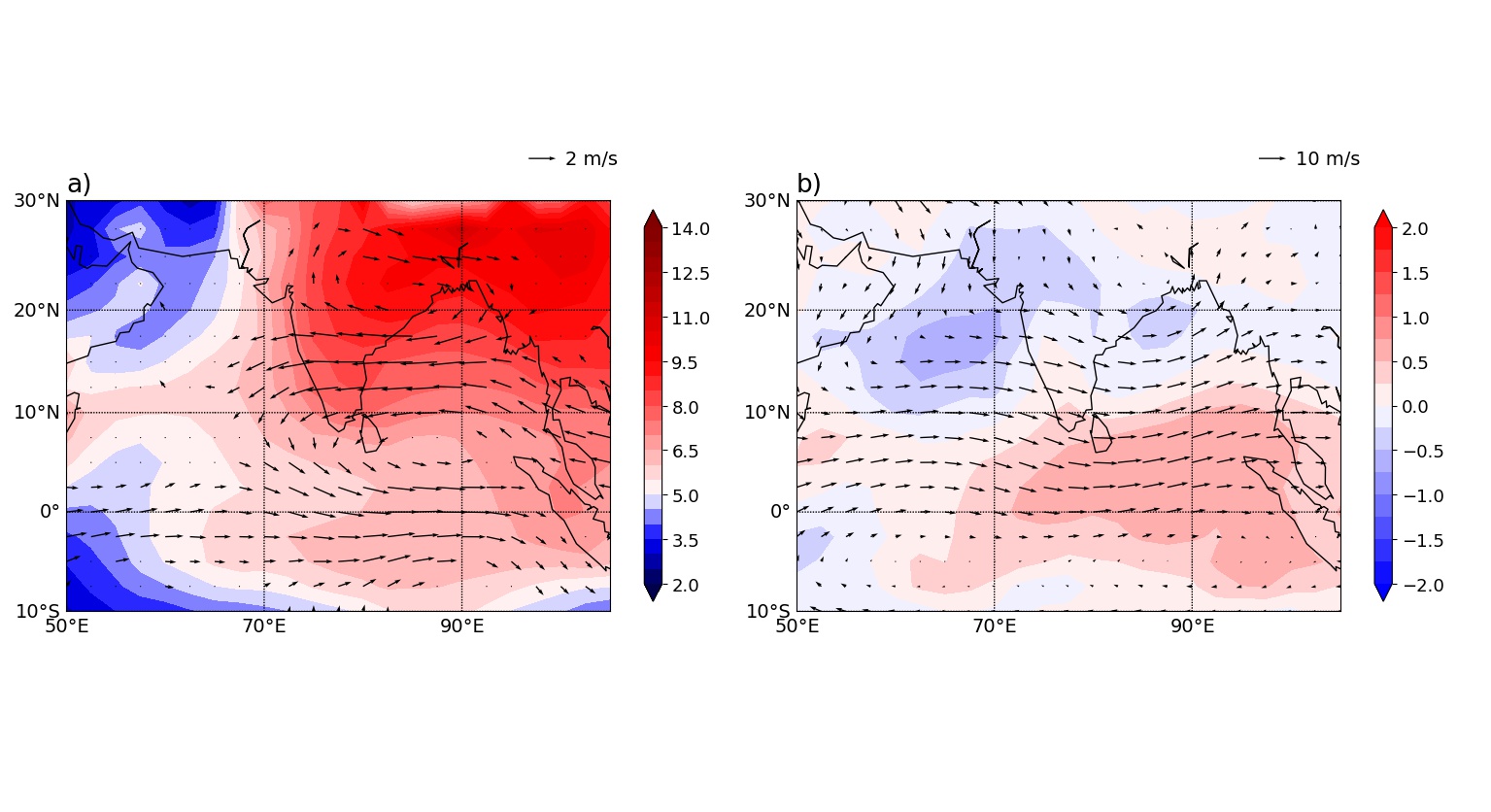}
    \caption{Same as Figure \ref{fig9}, but for Day $0$ of non-propagating composite.} 
    \label{fig10}
\end{figure}

\noindent To demonstrate how the BSISO wind anomalies (i.e., the eddies) advect background moisture and monsoon winds advect BSISO moisture anomalies in their respective places of dominance, we plot 700 hPa background moisture along with 700 hPa BSISO wind anomalies as well as 700 hPa BSISO moisture anomaly with background wind on Day -8 and Day 4 for the propagating composite in Figure \ref{fig9}. Similarly, Figure \ref{fig10} shows these fields on Day 0 for the non-propagating cases.

\noindent In the first stage for the propagating cases, when convection starts in the EIO, i.e., on Day -8, easterlies (which are a part of anticyclonic circulation associated with the suppressed convection on the land) are present from the equator to 15N, covering the AS and peninsular India (Figure \ref{fig9}a). 
These easterlies act upon the sharp zonally oriented gradient of background moisture over the AS (Figure \ref{fig9}a) and moisten this region. Interestingly over the BOB, the gradient of background moisture is more meridionally oriented, but the wind anomalies are north-easterly, so the advection of moisture is much weaker, thus the BSISO moisture anomaly (convection) moves much quicker over the AS than the BOB and eventually forms a slanted band. On Day -4 (Figure S4) we see a very similar picture with strong easterlies over the AS that cause horizontal advection of moisture into the AS sector, but these easterlies are not only associated with the anticyclonic circulation of the suppressed convection over the land, but also with the newly forming cyclonic Rossby response associated with the nascent convection over the EIO. This signal persists up to Day 0 and continues to moisten the AS. On the other hand, the negative BSISO moisture anomaly associated with the suppressed convection is stronger over the AS than over peninsular India, so background westerly monsoon winds cause dry advection near the coast and over peninsular India (Figure \ref{fig9}b). 

\noindent In the second stage, i.e., for northward movement over the SA, on Day 4 (Figure \ref{fig9}c), we see that the cyclonic Rossby gyre associated with the enhanced convection has a clear tilt from NW-SE (which is first visible on Day 4), and a well-formed vortex can be seen over the AS. Associated with this gyre, the wind anomalies over the land are south-easterly, and aligned with the background moisture gradient (the background moisture decreases from the head BOB-Bangladesh-Myanmar region towards the desert region in the North-West India/Pakistan). This wind taps into the moisture gradient and advects moisture from the BOB to northern India/Pakistan north of 20N. Further, the background monsoon wind, which is south-easterly over the BOB, acts upon the anomalous moisture gradient in the same direction (due to slanted convection band) and moistens the Bay (Figure \ref{fig9}d), thus the moisture anomaly and convection move further northward towards the north BOB. Similarly, monsoon winds advect anomalous moisture from AS towards the peninsular India. Note that, by Day 4 and Day 8 (Figure \ref{fig1}A and \ref{fig2}A), as the Rossby gyre is tilted and moves northward, the anomalous winds over the EIO become westerly and these act against the background moisture gradient to dry the region. As a result, the whole tilted band of anomalous moisture (convection) moves north from the EIO .     

\noindent For non-propagating events, we clearly see the Rossby gyre associated with the Gill-type response and prominent easterlies between 10-20N (Figure \ref{fig10}a), but it is comparatively weaker than the propagating cases, and most importantly, it is limited to the east of 70E, while the background moisture distribution is not very different. Even near 70E, the winds turn at the edge of the gyre, so they are more north-easterly/northerly than easterly. These wind anomalies can't successfully tap the moisture gradient present in the AS, and fail to amply moisten the region by advection. Moreover, due to the lack of strong moistening before Day 0, the negative moisture anomaly of the dry cycle over AS is quite strong, and the background westerlies act upon that to give rise to strong negative moisture advection near the coast (Figure \ref{fig10}b). Overall, unlike the propagating cases, sufficient moistening doesn't happen over the AS. At the next stage, the Rossby vortex doesn't tilt towards the north-west and the vast landmass north of 20N doesn't get moistened by it.    

\noindent Thus, in contrast to the suggestion by \cite{jiang2018unified}, strong easterlies over India don't guarantee robust northward propagation. In fact, we see the critical condition is that the easterlies need to extend far beyond 70E to amply moisten the AS. Why the strong moistening of AS is critical for the BSISO to reach further North into the South Asian land region? This question takes us back to the sudden tilt in the Rossby response (as seen between Day 0 to Day 8 of propagating cases), as the south-easterlies associated with the tilted Rossby response are the reason behind the moistening over most of the land region.

\section{Vorticity budget}

\noindent We now proceed to vorticity budget analysis with the aim of understanding the tilt of the Rossby response. The column-integrated version of the relevant equation reads \citep{wang2017quasi},
\begin{equation}
\langle\frac{\partial \zeta}{\partial t} \rangle'= \langle(-\omega\frac{\partial \zeta}{\partial p})\rangle'+ \langle(-\textbf{V}.\nabla_{h} \zeta)\rangle'+\langle(-v\frac{\partial f}{\partial y})\rangle'+ \langle-(\zeta+f)D]\rangle'+ \langle T_1\rangle' - \langle T_2\rangle' + \textrm{residual}, 
\label{e7}
\end{equation}
where, $\zeta=(\frac{\partial v}{\partial x}-\frac{\partial u}{\partial y})$ and $D=(\frac{\partial u}{\partial x}+\frac{\partial v}{\partial y})$ are the relative vorticity and divergence, respectively.  $\textbf{V}=u\textbf{i}+v\textbf{j}$  is the horizontal wind,  ${\nabla_{h}=\textbf{i}(\frac{\partial }{\partial x})+\textbf{j}(\frac{\partial }{\partial y}})$ is the horizontal gradient operator, $f$ is Coriolis parameter and $\omega$ is the vertical velocity in pressure co-ordinates. Prime denotes a 25-70 day anomaly as defined earlier. $T_1$ is given by $(\frac{\partial \omega}{\partial y})(\frac{\partial u}{\partial p})$ and $T_2$ is given by  $(\frac{\partial \omega}{\partial x})(\frac{\partial v}{\partial p})$, and $(T_1-T_2)$  is the tilting term. In this analysis, we separately show these first and second terms of the tilting, as there are conflicting views regarding which term is important in the BSISO.  $[-(\zeta+f)D]$ represents the stretching term, $\frac{\partial \zeta}{\partial t}$ is the local tendency of the relative vorticity, $(-\textbf{V}.\nabla_{h} \zeta)$ and $(-\omega\frac{\partial \zeta}{\partial p})$ represent the horizontal and vertical advection of relative vorticity, respectively, and $ (-v\frac{\partial f}{\partial y})$ is the vorticity generation due to the $\beta$ effect. 
Angle bracket here represents mass-weighted vertical integral in the lower free troposphere (850 to 600 hPa).

\begin{figure}
    \centering
    \includegraphics[width=\textwidth]{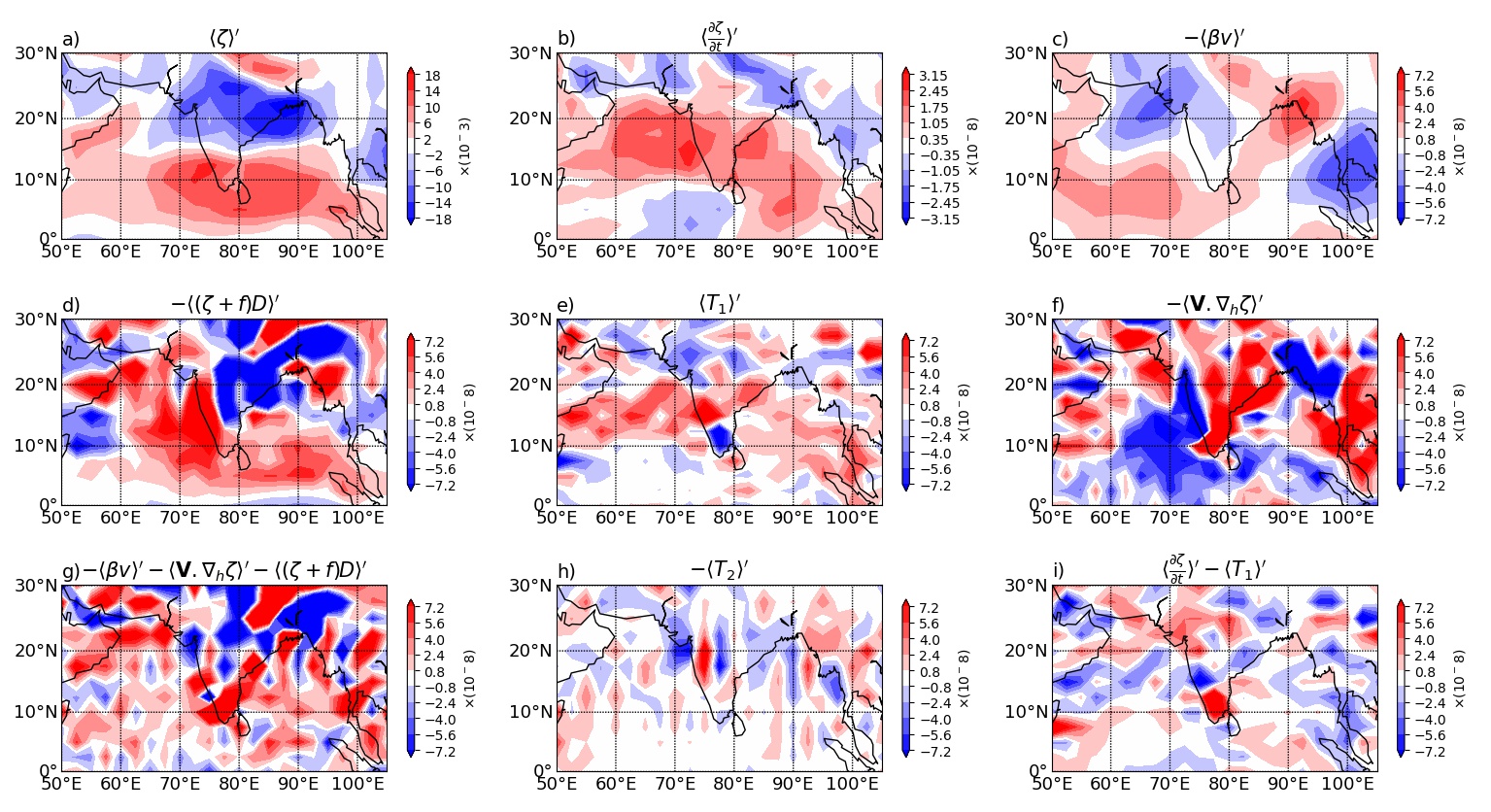}
    
    \caption{600-850 hPa (lower free-troposphere) integrated (a) Vorticity anomaly and dominant vorticity budget terms  as shown in Equation \ref{e7} and their combinations, namely, (b) Vorticity tendency, (c) $\beta$ term, (d) Vortex stretching, (e) First term of vortex tilting, (f) Horizontal advection, (g) Sum of $\beta$ term, vortex stretching and horizontal advection, (h) Second term of vortex tilting, (i) Vorticity tendency minus first term of vortex tilting on Day $0$ of the propagating composite. The unit of column integrated vorticity is kgm$^{-2}$ s$^{-1}$ and of budget terms are kgm$^{-2}$s$^{-2}$.}  
    \label{fig11}
\end{figure}

\noindent Figure \ref{fig11} shows the terms comprising the lower free-tropospheric vorticity budget and their combinations on Day 0 of the propagating cases. We chose Day 0 because, from Day 0 to Day 4, we see the abrupt northward jump of the Rossby gyre, with a clear tilt from the NW to the SE (Figure \ref{fig1}). Indeed, the Rossby gyre was zonally oriented on Day 0 (Figure \ref{fig1} and Figure \ref{fig11}a). In other words, positive vorticity anomalies moved northward much faster in the AS than in the BOB. 
In Figure \ref{fig11}a, we see zonally oriented cyclonic vorticity (associated with the Rossby response of the Equatorial convection) up to 15N. The tendency (Figure \ref{fig11}b) has a clear NW-SE tilt, which is expected as discussed above. Clearly, the prime contributor to this NW-SE tilted tendency is $T_1'$ (the component of the tilting term associated with the meridional gradient of anomalous vertical velocity), shown in Figure \ref{fig11}e. The tendency as well as $T_1'$ are particularly strong over the AS, compared to the BOB, which causes the generation of cyclonic vorticity over the AS on Day 4 and Day 8 and moves the vortex northward much faster over the basin. 

\noindent Interestingly, while $T_1'$ is the main contributor to the tendency, it is not the largest term in the budget. The largest terms in this budget are the stretching (Figure \ref{fig11}d) and the horizontal advection (Figure \ref{fig11}f), but they mostly cancel each other. The cancellation of these two terms is a feature noted in various instances, such as the global interaction of transient eddies and the seasonal mean flow \citep{HS} to synoptic systems in the Indian monsoon \citep{boos2015adiabatic,kushwaha2023}. Indeed, stretching, horizontal advection, and $\beta$ terms together are close to zero over the AS (Figure \ref{fig11}g), where the tendency is the strongest. Vertical advection (not shown) is negligible over the AS and BOB but has a small positive contribution over the land. As seen in Figure \ref{fig11}h, $-T_2'$ is also negligible, and is not important in BSISO propagation as was suspected earlier \citep{karmakar2022northward}. The residual (not shown) is non-negligible, but it is weak and negative over the region of strongest positive tendency over the AS and BOB, so it doesn't jeopardize our understanding. Above all, the contribution from all the terms except $T_1'$ is shown at the bottom right panel (Figure \ref{fig11}i; this includes the terms not explicitly shown in the figure) yields negative values over the region of positive tendency, but  $T_1'$ is much larger than the negative contribution from all other terms, thus resulting in a net positive vorticity tendency. In essence, we can conclude that  $T_1'$ is the term that dictates the NW-SE slanting of the Rossby gyre in the propagating composite. Near the equator, horizontal advection contributes to a negative tendency, so the whole gyre moves north. A similar process also holds on Day 4 (not shown), which tilts the gyre even more and generates stronger south-easterlies over land. In fact, the importance of the tilting term in northward propagation has been recently noted in other studies too \citep{li2021role,karmakar2022northward}.

\begin{figure}
    \centering
    \includegraphics[width=\textwidth]{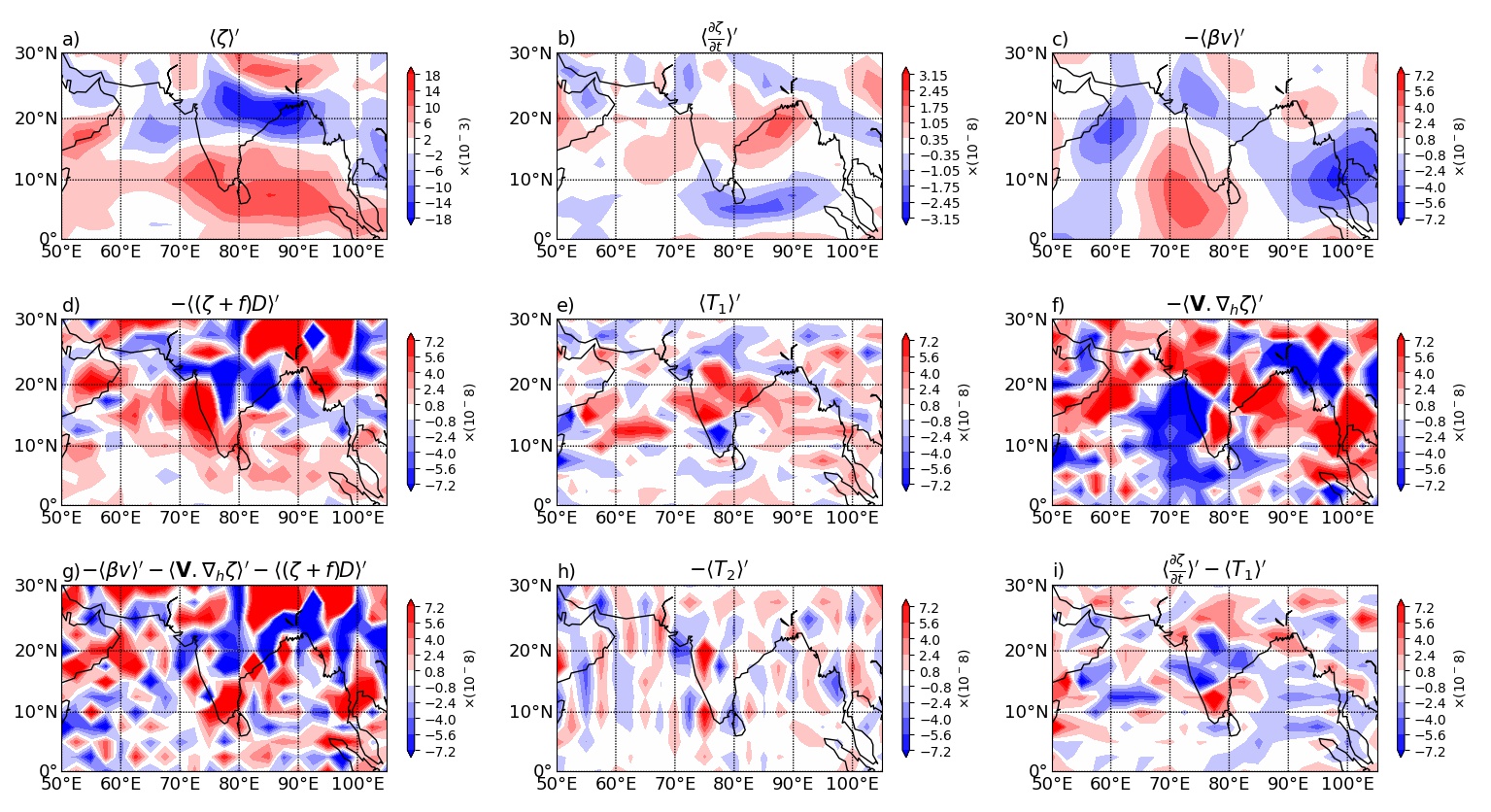}
    \caption{Same as Figure \ref{fig11}, but for Day $4$ of non-propagating composite.} 
    \label{fig12}
\end{figure}

\noindent With regard to the Rossby gyre in the non-propagating events, we focus on the Day 4 vorticity budget shown in Figure \ref{fig12}. We chose this day because, on Day 4, a clear well-formed Rossby signal is evident over EIO and Southern India, but it fails to propagate northward and weakens by  Day 8 (Figure \ref{fig2}B). As seen in Figure \ref{fig12}e, $T_1'$ is almost absent over the northern AS, there is a small positive patch below 15N, and the tendency term (Figure \ref{fig12}b) reflects a similar pattern. As in the propagating events, a cancellation between stretching (Figure \ref{fig12}d) and horizontal advection (Figure \ref{fig12}f) is present, and when added with the $\beta$ term (Figure \ref{fig12}c), their cumulative contribution over the AS is again close to zero (Figure \ref{fig12}g). 
Comparing with the propagating cases, we can conclude that the vortex doesn't propagate into the AS, and fails to generate the NW-SE tilted Rossby gyre due to a non-existent $T_1'$ term in the Northern AS. Over the Indian landmass, we see a contribution from $T_1'$, but it is not strong enough to amply counteract the opposing effects by other terms, though it does generate a small positive tendency, thus the gyre moves slightly northward over land, but a lack of tilt implies that it can't generate the required south-easterly winds. Moreover, near the equator, horizontal advection primarily contributes to the negative tendency to weaken the cyclonic vortex.

\subsection{Process of tilting}

\noindent The vorticity budget suggests that the question of northward propagation of BSISO over land boils down to explaining why a strong tilting term ($T_1'$) is generated over the northern AS region in case of the  propagating BSISO events once the Rossby signal associated with the nascent convection in the EIO region gets firmly established, while this doesn't happen for the non-propagating cases. To understand this in more detail, we decompose $T_1'$ into components comprising background and BSISO-related anomaly fields. Here, we only show ${(\frac{\partial \omega'}{\partial y})(\frac{\partial \Bar{u}}{\partial p})}$, which is the dominant term controlling $T_1'$. Physically, this term comprises the meridional gradient of anomalous vertical velocity and vertical shear of background zonal wind.
 
\begin{figure}
    \centering
    \includegraphics[width=\textwidth]{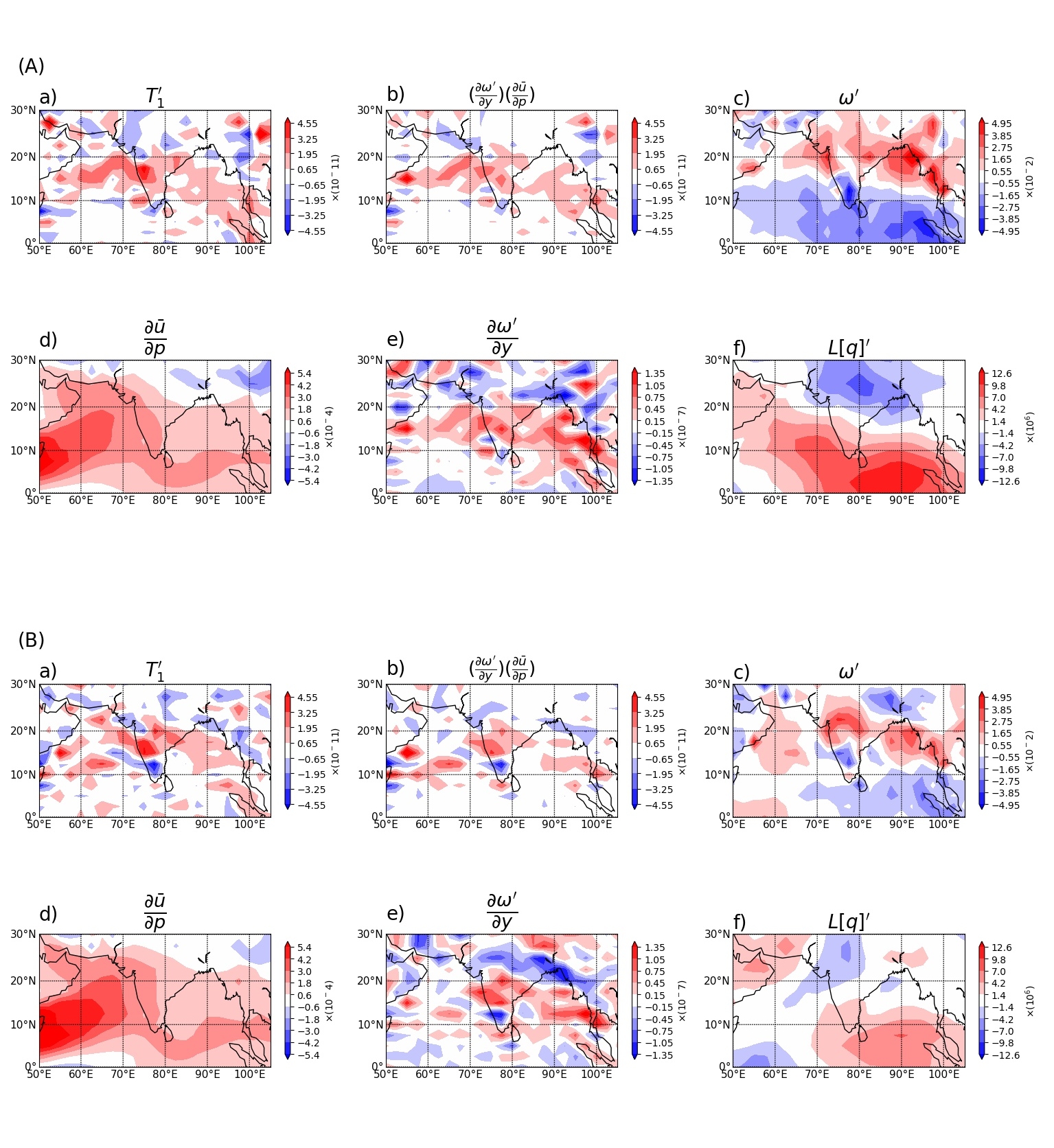}
    
    \caption{(A) $T_1'$  and important terms associated with $T_1'$ on Day 0 at 700 hPa for propagating cases, namely (a) First term (Dominant part) of vortex tilting anomaly $T_1'$, (b) Its dominant linearized component, (c) Anomalous vertical velocity, (d) Vertical shear of background zonal wind, (e) Meridional gradient of anomalous vertical velocity at 700 hPa and (e) Column-integrated specific humidity anomaly. Units are s$^{-2}$, s$^{-2}$, Pa s$^{-1}$, m Pa$^{-1}$
    s$^{-1}$, Pa (ms)$^{-1}$ and J m$^{-2}$ respectively. (B) Same as (A), but for Day 4 of non-propagating composite.}
    \label{fig13}
\end{figure}

\noindent To investigate the $T_1'$ term, we again focus on a single level (700 hPa), which is a representative level for the lower free-troposphere, and was also the level chosen for understanding the moistening process. In Figures \ref{fig13}(A) and \ref{fig13}(B), we have shown $T_1'$ and its dominant components and other relevant fields at 700 hPa for propagating and non-propagating cases respectively. Examining Figure \ref{fig13}, we see most of $T_1'$ is mostly captured by  ${(\frac{\partial \omega'}{\partial y})(\frac{\partial \Bar{u}}{\partial p})}$ (Figure \ref{fig13}(A)b and Figure \ref{fig13}(B)b) for both the propagating and non-propagating cases. 
The vertical wind shear, Figure \ref{fig13}(A)d and Figure \ref{fig13}(B)d), is very similar for both cases, and clearly, the difference between the two types of events arises from  ${(\frac{\partial \omega'}{\partial y})}$(\ref{fig13}(A)e and \ref{fig13}(B)e). Like $T_1'$, this term is positive north of the convection, tilted from NW-SE and very strong over the northern AS for the propagating cases, while for the non-propagating cases, it is limited to the land and the BOB, and essentially non-existent over the AS. The vertical shear of background zonal wind (Figure \ref{fig13}(A)d) is also very large over the northern AS, compared to the BOB so it amplifies  $T_1'$ over this region for the propagating cases in Figure \ref{fig13}(A)b. Note that, ${(\frac{\partial \omega'}{\partial y})}$ is also strong over the BOB, but a weaker background shear renders tilting to be comparatively weak in this region. 

\noindent The importance of vorticity generation by the tilting term over the northern AS needs to be especially highlighted, as strong tilting over the northern AS causes the NW-SE tilted vortex structure, that is elongated from the BOB almost up to the Arabian peninsula, which generates south-easterlies and is very important for moistening over land. Previous studies that have studied tilting or vertical shear have not stressed the importance of this term in this area, and have mainly focused their attention on either the BOB, or the region south of 15N \citep{neena2017model,li2021role,li2022key}, thus missed the importance of northern AS in generating the elongated and tilted vortex that is crucial for 
the northward propagation of convection over land. Moreover, vertical shear of zonal wind as well as tilting was calculated in the whole free-troposphere (850-200 hPa), and this is much weaker over northern AS compared to most of the BOB or the entire area south of 15N. When we focus our attention on the lower free-troposphere, which is most important for the moist processes, we found that vertical shear over the northern AS is comparable to the value over southern AS, and stronger than the BOB.

\noindent Delving deeper, to explain why ${(\frac{\partial \omega'}{\partial y})}$ is strong over the northern AS for propagating cases, but not so for the non-propagating cases, we shift our focus to the orientation of $\omega'$ (Figure \ref{fig13}(A)c and Figure \ref{fig13}(B)c). For the propagating cases, by Day 0, anomalous negative values associated with ascending motion engulf all the EIO as well as the southern AS, while for the non-propagating cases, even on Day 4, this is limited to the EIO itself (and to some extent peninsular India), with much weaker magnitude, and there is no ascending motion over the AS. Moreover, over the AS, for the propagating cases, descending motion is confined near 20N, but for the non-propagating cases, weak anomalous descending motion is visible up to the southern AS. 
Thus, a strong meridional gradient of anomalous vertical velocity exists over the northern AS region for the propagating cases(which generates strong tilting as shown), while it doesn't for the non-propagating cases. The distribution of $\omega'$ closely follows the distribution of column integrated moisture anomalies (Figure \ref{fig13}(A)f and Figure \ref{fig13}(B)f) for both the composites, as in the `moisture mode' framework, anomalous moisture causes the convection that manifests in anomalous ascending/descending motion. Though $\omega'$ and column-integrated moisture anomaly for propagating cases are highly coherent over the EIO and southern AS, in the northern AS, moisture has a slight lead, which again hints that some moisture preconditioning might be necessary for this region to generate convection.  

\noindent Overall, the generation of the NW-SE tilted elongated band of ${(\frac{\partial \omega'}{\partial y})}$ for propagating cases, with large values in the northern AS, and its absence in non-propagating cases is dictated by the presence/absence of the strong moisture(convection) anomalies over the southern AS. For the propagating cases, when convection matures in the EIO, easterlies to the north of convection advect significant moisture into the southern AS, which results in convection and ascending motion. But, moistening does not occur over the BOB at this stage. Thus, the slanted shape of convection and ascending motion, behind the anomalous suppressed convection and associated descent, in presence of vertical shear in background zonal wind gives rise to the NW-SE slanted tilting term to the north of the existing convection. For the non-propagating cases, weak easterlies can't moisten the southern AS, thus convection and associated ascent is absent in the AS. Together with a weakening descending motion over the AS, this results in the absence of strong meridional gradient in anomalous vertical velocity, particularly over the northern AS. Hence, vortex tilting is not generated in the northern AS, and the vortex doesn't move northward with the NW-SE tilted structure necessary for subsequent moistening over land. Indeed, this clearly illustrates the coupling between moisture and vorticity dynamics for the propagation of the BSISO.  
\section{Discussion}

\noindent A leading proposal for the northward movement of the BSISO involves moistening to the north of the existing convection via boundary layer moisture convergence (BLMC), which is caused by the generation of barotropic vorticity in the free troposphere ahead of the convection \citep{jiang2004structures,drbohlav2005mechanism,bellon2008instability,dixit2011role}. 
Theories that follow this framework agree on the generation of vorticity to the north of the convection, but how the vorticity is generated differs from theory to theory \citep{bellon2023selected}. In a few idealized models, the primary factor responsible for the generation of barotropic vorticity is vertical shear of the background zonal flow which couples baroclinic and barotropic modes \citep{jiang2004structures,drbohlav2005mechanism}. 
Another possibility is the 
generation of free-tropospheric vorticity north of the existing convection due to advection of vorticity by the vertically sheared meridional mean flow \citep{bellon2008instability}. 
In another view, $\beta$ drift is important \citep{boos2010mechanisms}, but again frictional convergence due to boundary layer dynamics is the key. Similarly, the convective momentum transfer mechanism \citep{kang2010mechanism} also requires moistening via moisture convergence, with an essential role for the vertical shear of the background flow.  
Here, we have demonstrated that vorticity generation due to tilting caused by the vertically sheared background zonal wind is an integral part of northward propagation. But, boundary layer dynamics don't play a key role in propagation, and moistening doesn't happen via BLMC, rather, BLMC has a negative impact on the moistening in front of the convection. 
In this regard, the BSISO is different from the MJO, which is driven eastward by both BLMC as well as free-tropospheric horizontal moisture advection \citep{hsu2012role,adames2015three}.

\noindent In this study, we have shown that horizontal moisture advection, particularly in the free troposphere, not in the boundary layer as suggested by few \citep{jiang2004structures,demott2013northward}, is the primary driver behind the northward propagation of the BSISO. Despite growing consensus about the importance of horizontal moisture advection, there is a debate about the exact processes involved \citep{kikuchi2021boreal}. These discrepancies have arisen because moisture budgets have been performed over different regions \citep{kikuchi2021boreal}; indeed, we have clearly demonstrated that differing processes are in action in different regions and stages of the BSISO. Thus, a convection-centric composite view taken by previous studies \citep{jiang2004structures,abhik2013possible} might not be the best way to look at these processes. From our analysis, background moisture advection by BSISO winds is dominant over the AS, but background wind advecting BSISO moisture anomalies is more important over the BOB, and also, this gains prominence slightly later in the BSISO life-cycle than the moistening over the AS. 
The studies that primarily focused on the AS or a larger domain comprising both the AS and BOB, and also on the early stages of propagation, found the first process to be more important \citep{jiang2018unified} because of its stronger magnitude over the AS at the early stages, while the studies primarily focused over the BOB found the later to be important \citep{ajayamohan2011poleward,sooraj2013boreal,wang2020diagnosing}. To the best of our knowledge, the importance of both BSISO wind advection of background moisture and mean flow advection of BSISO anomalous moisture has only been noted by a handful of studies \citep{adames2016seasonality,gao2019diagnosing,karmakar2020differences,chen2021diversity}, and our results confirm and elucidate that finding. Equally important are the moistening processes over land, both in the peninsular region and poleward of 20N. 
We show that north of 20N, anomalous BSISO south-easterlies (north-westerlies) tap in the north-westward gradient of background moisture and moisten (dry) this region by horizontal advection, while, in agreement with \cite{jiang2018unified}, the peninsular Indian region is moistened via ``column-process". 

\noindent Noting the importance of background moisture advection by the BSISO horizontal winds, \cite{jiang2018unified} proposed that the mean moisture pattern during the boreal summer largely shapes the propagation of the BSISO, similar to its wintertime counterpart MJO, and both BSISO and MJO can be explained by an unified moisture mode framework --- in fact, the model for the BSISO and MJO by \cite{wang2022unified} follows such a prescription. 
While we confirm that the mean moisture distribution is essential in the propagation of the BSISO, this alone can't explain all of its defining features, as we have shown that the monsoon background flow is of critical importance, both due to the role of the vertical shear in generating the vortex tilting (which in turn facilitates moisture advection beyond 20N) and the role of the mean flow in the advection of moisture anomalies over the BOB. 
Thus, a minimal model of the BSISO in this framework should include a background wind with realistic vertical shear, moisture gradients like those over the AS, BOB and SA with active water vapor dynamics, as well as prognostic momentum equations that interact with the moisture via convection. 

\noindent  Our results also have possible theoretical implications outside the context of BSISO. \cite{sobel2001weak} have shown that under the weak temperature gradient approximation, moisture can drive large scale rotational flow by generating stretching and horizontal advection through divergence, and this is regarded as a cornerstone of coupling between moisture (convection) and circulation. Similarly, \cite{rydbeck2015convective} observed that moisture generated stretching influences the circulation for easterly waves, and this mechanism has also been discussed in the context of few simplified MJO models \citep{Ahmed2021,hayashi2017new}. Here we find that, with a suitable background flow, moisture can lead to a large scale rotational flow by generating vortex tilting, and this understanding can be incorporated in future theoretical models of convectively coupled systems.

\noindent As far as air-sea interaction is concerned, it can help in the propagation of convection via enhanced moistening by two possible  ways \citep{kikuchi2021boreal}. SST anomaly gradients can induce BLMC, which is proposed to be effective in case of MJO \citep{hsu2012role,de2020atmospheric}, the other being the enhancement of surface evaporation. We have shown that both of these terms in the moisture budget oppose moistening to the north of the convection. Thus, air-sea interactions at intraseasonal scales do not appear to have crucial role in driving the convection northward (see also  \cite{gao2019diagnosing}). Though, increased SST in front of convection might help BSISO indirectly; specifically, without the intraseasonal fluctuations of SST, the drying terms to the north of the convection (evaporation and vertical advection) might have been even larger and thus could retard the progress of the BSISO \citep{bellon2008ocean,gao2019diagnosing}.

\begin{figure}
    \centering
    \includegraphics[width=16 cm]{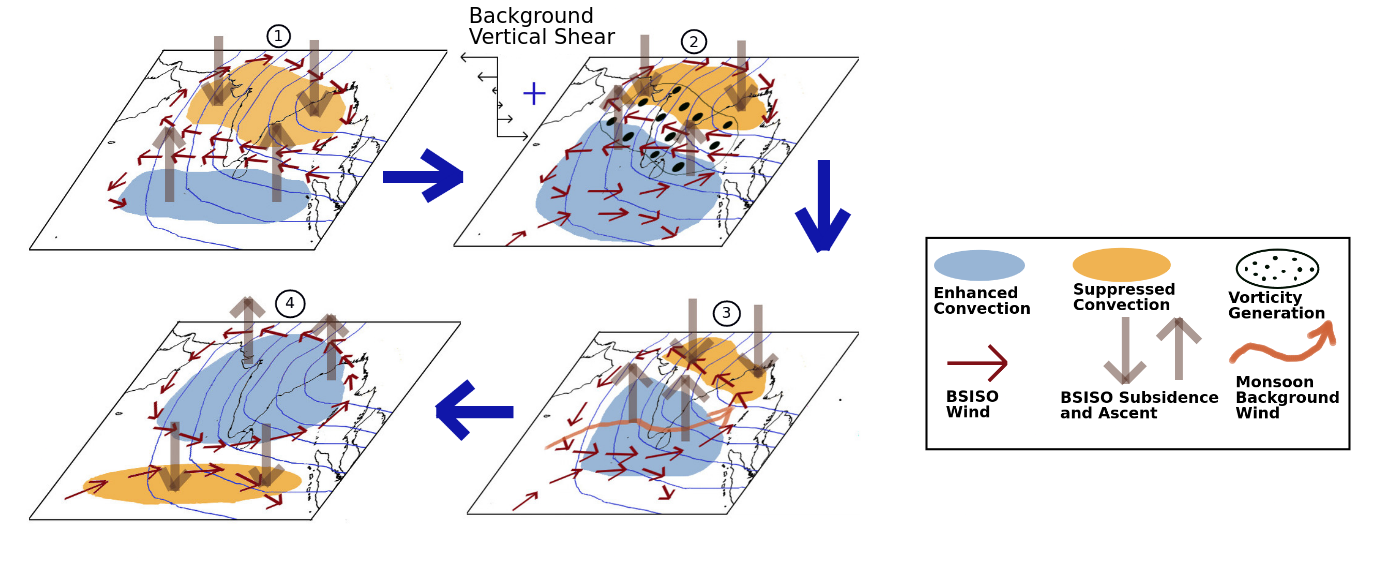}
    
    \caption{Schematic diagram depicting the mechanism of northward propagation of convection from the Equatorial Indian Ocean. Blue contours represents seasonal specific humidity (moisture) distribution. All other symbols are described in legend box in the right.}
    \label{fig14}
\end{figure}

\section{Summary and Conclusions} 

\noindent In this study, we have investigated the mechanisms behind the northward propagation of BSISO over South Asia using moisture and vorticity dynamics. We have identified two types of BSISO events, one propagates northward to the South Asian landmass from the Equatorial Indian Ocean (EIO), while the other doesn't. Comparing their propagation dynamics, we were able to identify the critical mechanisms behind northward propagation.

\noindent We confirmed that BSISO convection anomalies are generally collocated with the column-integrated moisture anomalies. Moisture and vorticity budgets 
showed that northward propagation of BSISO convection involves a chain of dynamically as well as thermodynamically connected processes. For propagating cases, easterlies on the southern flank of the anticyclonic Rossby gyre associated with the previous cycle of suppressed convection, as well as the easterlies in the northern flank of the cyclonic Rossby gyre associated with the new area of enhanced convection over the EIO engulf most of the Arabian Sea (AS). This aligns with the background gradient of moisture and moistens the atmosphere over the southern Arabian Sea by free-tropospheric horizontal advection. Over the Bay of Bengal (BOB), the background moisture gradient is primarily meridional while BSISO winds are north-easterly, thus the moistening due to advection is weaker. `Column-process' acts against this moistening, but the advection is stronger in the AS and hence convection rapidly enters the southern AS from the EIO resulting in the initial NW-SE tilted band of convection associated with the BSISO. 

\noindent At this stage, as a tilted belt of strong convection is present behind the area of suppressed convection associated with subsidence, a similarly tilted belt of a meridional gradient of anomalous vertical velocity associated with the anomalous convection comes into being from the northern AS to the southern BOB. The vertical easterly shear in the background monsoon wind acts upon this gradient to generate an NW-SE slanted vortex tilting term, which dominates the vorticity tendency and leads to the formation of a slanted vorticity anomaly north of the existing convection. Thus, the cyclonic Rossby gyre associated with the enhanced BSISO convection acquires a clear NW-SE tilt while moving northward from the EIO. On the northern flank of the tilted Rossby gyre, anomalous south-easterlies advect the background moisture from the moisture-rich northern BOB/Bangladesh region in the southeast towards the drier region in the northwest. Thus, again, the anomalous BSISO winds tap the background moisture gradient and moisten the vast expanse of land north of 20N, and convection marches further northward. On the southern flank of this Rossby gyre, westerlies advect dry air over the EIO, thus convection dies down, and the entire band of convection shifts from the equatorial region to the north. Notably, over the BOB, the moistening process is different. Once the tilted structure of convection comes into being, the background south-easterly monsoon wind acts on the anomalous moisture gradient to help initiate convection over the northern BOB. These chain of processes have been shown by a schematic diagram in Figure \ref{fig14}.

\noindent For the non-propagating BSISO cases, as nascent convection starts to gain strength over the EIO, similar to the propagating cases, a prominent Rossby response to the convection gets generated, but the easterlies associated with the Rossby gyre are primarily confined to the BOB and southern India. The anti-cyclonic wind response associated with the previous cycle of suppressed convection is also much weaker in strength. These relatively weaker winds over the Arabian Sea don't moisten the region over the southern AS to initiate convection, despite the presence of a prominent zonal gradient of background moisture. Here the chain of processes breaks, in the absence of strong convection behind the zone of subsidence associated with suppressed convection, the meridional gradient of vertical velocity is very weak, particularly over the northern AS. Hence, in spite of the presence of strong easterly vertical shear of background zonal wind, a strong vortex tilting term doesn't get generated, and the slanting of Rossby gyre doesn't occur. As a result, the convection stalls over the EIO, and the westerlies in the southern flank of the Rossby gyre of the enhanced convective signal eventually kill the convection.



\noindent Overall, the northward propagation of BSISO can be understood as a horizontal advection driven moisture mode acting under the influence of the background vertical shear of the zonal monsoon wind. It is a classic case of convectively coupled dynamics, where moisture and circulation influence each other to dictate the propagation of convection. Thus, the BSISO is a dynamic moisture mode, where active feedback between moisture and dynamics is essential. 
Indeed, vertical shear of the background zonal wind and the zonal gradient of background moisture over the AS are necessary conditions for northward propagation, but they are not sufficient by themselves. The critical difference between the propagating and non-propagating cases arises from the strength and extent of anomalous easterlies over the southern AS when convection starts over the EIO, as moistening over the southern AS by these easterly winds is crucial for the northward propagation of the BSISO. Moreover, moistening primarily occurs above the boundary layer, in the lower free-troposphere. On the other hand, only moisture advection, without the role played by vertical shear, can generate limited northward propagation, up to the AS, but it can't explain the moistening process over land beyond 20N. It is also unable to explain the observed tilt in the Rossby gyre. Till date, the vortex tilting mechanism via vertical shear and horizontal advection mechanism were thought as either somewhat contradictory \citep{yang2019mechanisms}, or independent \citep{li2021role,wang2022unified}. Our results show that, in fact, vortex tilting and the moisture advection processes can't be looked at separately, they work hand in hand and influence each other to facilitate the large-scale northward propagation of the BSISO from the equatorial Indian Ocean to the South Asian landmass.

\section{Open Research}
\noindent All data used in this study is open access. NOAA OLR data can be accessed from \url{https://psl.noaa.gov/data/gridded/data.interp_OLR.html}, ERA5 data can be accessed via \url{https://cds.climate.copernicus.eu/cdsapp/dataset/reanalysis-era5-pressure-levels},  Details for accessing and using the Windspharm package \cite{dawson} are given at \url{https://ajdawson.github.io/windspharm/latest/}. 

\noindent {\bf Acknowledgements:} SG acknowledges financial assistance from Council for Scientific and Industrial Research (CSIR) and the Divecha Centre for Climate Change, IISc. SG thanks ABS Thakur, Soumalya Seal and Soham Halder for their help in coding and Shubhrangshu Biswas for his help in making the schematic. SG also thanks Prof. J Srinivasan and Prof. Debasis Sengupta for their insightful comments.


\bibliographystyle{apalike}
\bibliography{example.bib}

    \end{document}